\documentclass[aps,pra,twocolumn,groupedaddress]{revtex4-1}
\usepackage{here}    % comment out for preprint style 
\usepackage{graphicx,amsmath,bm}
\begin{document}
\title[]{Synthetic gauge field and pseudospin-orbit interaction in a stacked two-dimensional ring-network lattice}
\author{Tetsuyuki Ochiai}
\affiliation{Research Center for Functional Materials, National Institute for Materials Science (NIMS), Tsukuba 305-0044, Japan}
\date{\today}

\begin{abstract}
We study the effects of a synthetic gauge field and pseudospin-orbit interaction in a stacked two-dimensional ring-network model. 
The model was introduced to simulate light propagation in the corresponding ring-resonator lattice, and is thus completely bosonic. 
Without these two items, the model exhibits Floquet-Weyl and Floquet-topological-insulator phases with topologically gapless and gapped band structures, respectively.  
The synthetic magnetic field implemented in the model results in a three-dimensional Hofstadter-butterfly-type spectrum in a photonic platform. The resulting gaps are characterization by the winding number of relevant S-matrices together with the Chern number of the bulk bands. The pseudospin-orbit interaction is defined as the mixing term between two pseudospin degrees of freedom in the rings, namely, the clockwise and counter-clockwise modes. 
It destroys the Floquet-topological-insulator phases, while the Floquet-Weyl phase with multiple Weyl points can be preserved by breaking the space-inversion symmetry. Implementing both the synthetic gauge field and pseudospin-orbit interaction requires a certain nonreciprocity. 

\end{abstract}
%---- APS --------------------------
\pacs{03.65.Vf, 73.20.-r, 71.70.Ej}
\keywords{}
\maketitle
%--- IOP ---------------------------

\section{Introduction}
Recently, much attention has been paid to  two-dimensional (2D) optical ring-resonator lattices  as a platform for topological photonics, synthetic gauge field, and many-body physics of photons \cite{hafezi2011robust,hafezi2013imaging,hafezi2013non,PhysRevLett.110.203904,PhysRevB.89.075113,lu2014topological}. 
Many interesting phenomena such as the Hofstadter butterfly \cite{hafezi2011robust} and chiral edge states \cite{hafezi2011robust,hafezi2013imaging,PhysRevLett.110.203904,PhysRevB.89.075113}  can emerge in the systems, by their analogy to quantum Hall systems. 
Actually, the ring-resonator lattice can be viewed as an optical realization of the Chalker-Coddingnton (CC) network model \cite{0022-3719-21-14-008}, which was introduced to simulate the Anderson localization in quantum Hall systems.

The CC network model without disorder is an unconventional Floquet-Bloch system without any time-periodic drive \cite{PhysRevB.89.075113}. 
Conventional Floquet-Bloch systems involve a periodic drive such as by irradiating circular-polarized light, often resulting in topological phases \cite{PhysRevB.79.081406,PhysRevLett.105.017401,lindner2011floquet,wang2013observation,PhysRevB.89.121401}. 
The network model hosts the so-called anomalous Floquet insulator phase \cite{PhysRevLett.110.203904,PhysRevB.89.075113}, which has chiral edge states even if the Chern number is zero \cite{PhysRevB.82.235114,PhysRevX.3.031005}. Moreover, unpaired Dirac cone emerges  at the phase boundary between the anomalous Floquet insulator and normal insulator phases \cite{PhysRevB.54.8708,2040-8986-18-1-014001}\footnote{Systems without the time-reversal symmetry and space-inversion symmetry can exhibit unpaired Dirac cone \cite{haldane1988mqh,ochiai2009photonic}. However, the unpaired Dirac cone concerned here emerges in a system with the space-inversion symmetry but without the time-reversal symmetry.}. 
Such exotic behaviors of the model enable us to investigate its three-dimensional (3D) generalization.

In the previous works \cite{ochiai2015gapless,ochiai2016floquet}, the author studied 3D  generalizations of the CC network model, having their optical realization by ring resonators in mind. In Ref. \onlinecite{ochiai2016floquet}, we show that a multilayer stacking of the CC model with interlayer scattering channels exhibits 3D topological phases of gapless and gapped band structures. In particular, Weyl points emerge in a robust manner in certain parameter regions. There, the system is in the Floquet-Weyl (FW) phase.  The Weyl points disappear by a formation of line nodes and the system turns into either nontopological or topological gapped phases. We call these phases the normal insulator (NI) and Floquet topological insulator (FTI) phases, respectively.   Similar FW and FTI phases are obtained in a 3D waveguide network \cite{PhysRevB.93.144114}, which can be viewed as another 3D generalization of the CC network model. The FW phases are also obtained in 3D topological insulators \cite{wang2014floquet} and stacked graphene systems \cite{PhysRevB.93.205435} by irradiating circular-polarized light.

It was also pointed out in Ref. \onlinecite{ochiai2016floquet}, that the stacked 2D network model can host 3D synthetic gauge fields just by adjusting the structure. 
This implementation of the synthetic gauge field is completely different from that in cold-atom systems, where  rotation or laser-assisted tunneling scheme is employed \cite{jaksch2003creation,PhysRevLett.92.040404,lin2009synthetic,gerbier2010gauge}. 
The gauge fields are for photons in the corresponding ring-resonator lattice, although photons themselves are gauge bosons.  The model can also host naturally a pseudospin-orbit interaction (PSOI) by regarding the degenerate clockwise and counter-clockwise modes of the rings as a bosonic pseudospin. This property also forms a remarkable contrast to spin-orbit interactions in cold atom systems \cite{galitski2013spin}.
It is interesting how the synthetic gauge fields and PSOI affect the topological phases. This question is important not only from a fundamental-physics viewpoint but also from a practical viewpoint. How are the resulting surface states, which can be used for a novel optical waveguide,  robust against perturbations?

Such a question highlights distinct features of our bosonic model compared to conventional electronic systems. 
In electronic systems, it is well established that topological bands are accompanied by characteristic surface states \cite{wen1992theory,hatsugai1993cna}. For Weyl semimetals, surface states called Fermi arc emerge \cite{PhysRevB.83.205101,PhysRevLett.107.127205,PhysRevLett.107.186806,PhysRevA.85.033640,liu2016evolution}. Their dispersion curve connects two Weyl points (of opposite chiralities) projected on the surface Brillouin zone.  For topological insulators, gapless Dirac-cone surface states emerge. They are robust against various perturbations and are protected by the Kramers degeneracy due to the time-reversal symmetry (TRS) \cite{fu2007topological,xia2009observation,PhysRevLett.103.146401,hasan2010colloquium,RevModPhys.83.1057}. These properties are expected to share, in part, in our model. However, our system is completely bosonic, and can be viewed as a Floquet system.  These distinct features  may disturb to draw a simple analogy with conventional electronic systems.

In this paper, we investigate effects of the synthetic gauge fields and PSOI in the stacked 2D ring-network lattice. A synthetic magnetic field implemented  in the lattice yields a 3D analogue of the Hofstadter butterfly and nontrivial topology in the bulk band structure. The latter property results in emergence of gapless surface states that are not allowed in the system without the gauge field. A simple relation of the bulk-edge correspondence is also drawn. The PSOI is shown to break the FTI phase, whereas the FW phase can be preserved by breaking the space-inversion symmetry (SIS). We also consider the compatibility between the synthetic gauge field and PSOI.  We found that it requires a certain nonreciprocity.

This paper is organized as follows. In Sec. 2, we briefly overview the stacked 2D network model. In Sec. 3, synthetic gauge fields are implemented.  A 3D analogue of the Hofstadter butterfly is presented, and nontrivial topology due to the gauge fields is investigated. Section 4 is devoted to study effects of PSOI in the topological phases of the system without the PSOI. Finally in Sec. 5, we summarize the results.

\section{Quick overview of the network model}
 
\begin{figure}
\begin{center}
\includegraphics[width=0.35\textwidth]{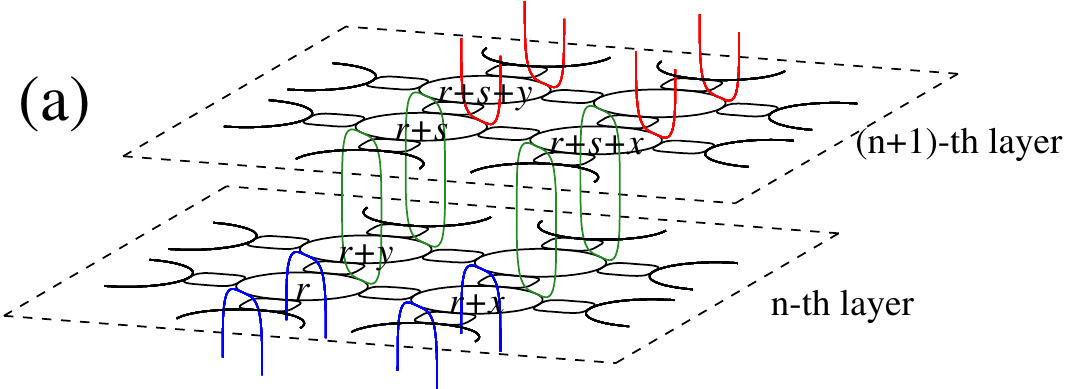}\\
\includegraphics[width=0.2\textwidth]{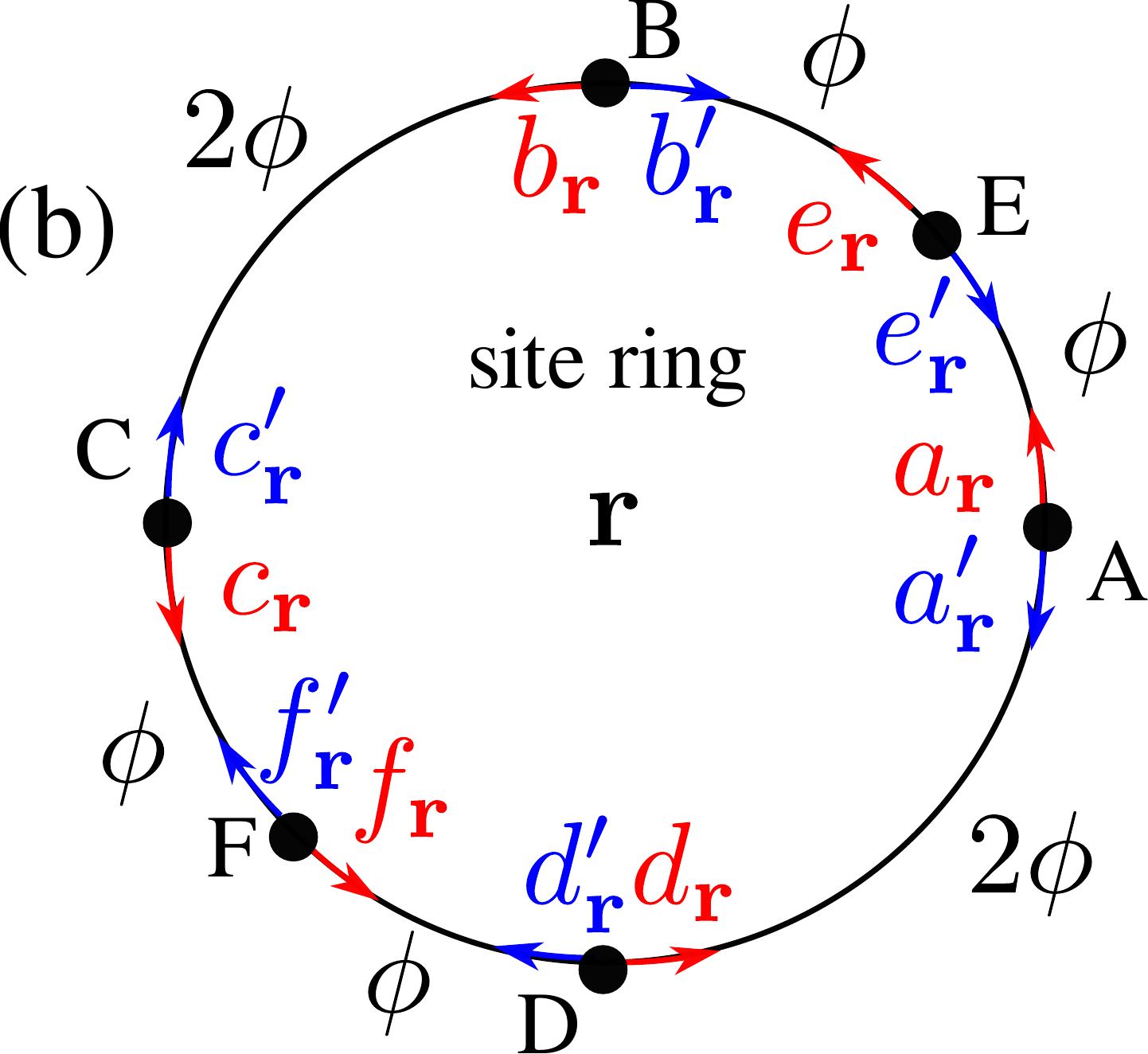}\\
\includegraphics[width=0.3\textwidth]{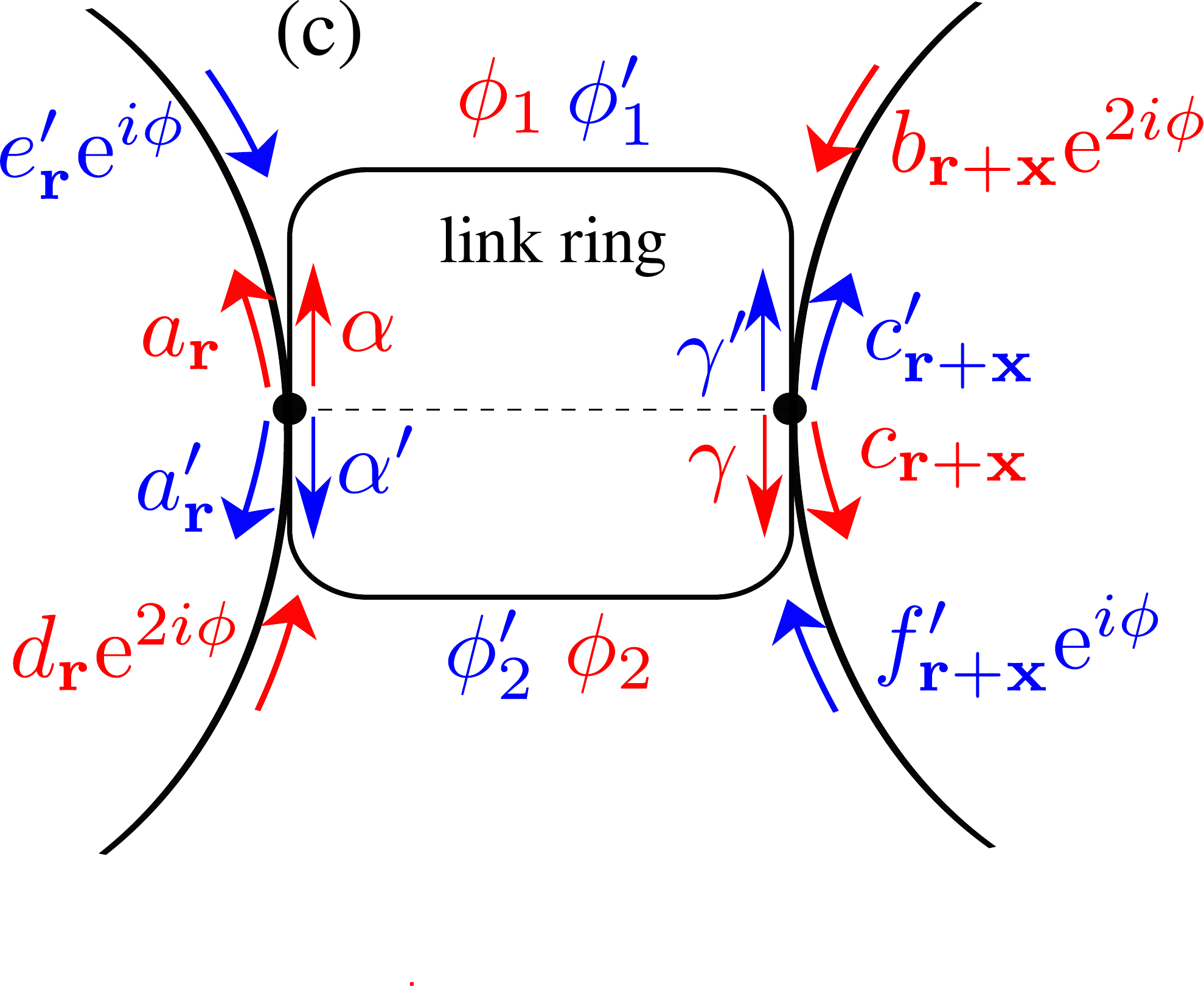}
\end{center}
\caption{\label{Fig_setup} (Color online) Schematic illustration of the stacked two-dimensional ring-network model under study. (a) The model consists of multilayer stacking of identical two-dimensional ring-network layers. Each layer of the Lieb lattice is composed of two types rings, site ring and link rings. The interlayer scattering takes places with vertical link rings shown in red, green, and blue colors for visibility. (b) Mode amplitudes are defined at nodes A, B, C, D, E, and F of the site ring. The mode acquires propagation phase $\phi$ for every one eighth propagation of the site ring. (c) Mode amplitudes relevant to the scattering at nodes A and C. The link ring is placed in between two adjacent site rings. The propagation phases of the clockwise (counter-clockwise) mode in the upper and lower parts of the link ring are denoted as $\phi_1(\phi'_1)$ and $\phi_2(\phi'_2)$, respectively. }
\end{figure}

The stacked 2D ring-network lattice proposed in Ref. \onlinecite{ochiai2016floquet} consists of the multilayer stacking of identical 2D ring-network layers with a lateral shift and interlayer scattering channels. For a concrete description of the model, please consult Ref. \onlinecite{ochiai2016floquet}. Figure \ref{Fig_setup} shows a schematic illustration of the system under study. 
%%%% Fig. 1 %%%%%%%%%%%%%%%%%%%%%%%%5
%%%%%%%%%%%%%%%%%%%%%%%%%%%%%%%%%%%%%%%
The relative shift  ${\bm s}$ between two adjacent layers is given by ${\bm s}=(\hat{x}+\hat{y}+\hat{z})/2$, where $\hat{\mu}$ ($\mu=x,y,z$) is the unit vector parallel to the $\mu$ direction. 
Each layer forms the Lieb lattice composed of two types of the rings,  site ring and link ring.  
In each site ring, we introduce mode amplitudes $\alpha_{\bm r}$ and $\alpha'_{\bm r}$ ($\alpha=a,b,\dots,f$) that correspond to counter-clockwise and clockwise propagation modes at the nodes A-F, respectively. These modes are scattered at the nodes, where the site rings are contact with the link rings. This scattering is described by the hopping S-matrices among the nearby site rings. These modes are degenerate in the site rings and thus acquire the same propagation phase $\phi$ as they travel one eighth of the site ring.

In this lattice structure, we can separate the clockwise modes of the site rings from the counter-clockwise ones by the directional coupling at the nodes. Namely, a destructive interference between the site and link rings  with the opposite propagation directions around the nodes prohibits the mixing among red and blue flows in Fig. \ref{Fig_setup}. Therefore, if we focus on the counter-clockwise modes of the site rings, the S-matrices are written as 
\begin{align}
&\left(\begin{array}{c}
a_{\bm r}\\
c_{{\bm r}+\hat{x}}
\end{array}\right)=S_1 \left(\begin{array}{c}
d_{\bm r}\\
b_{{\bm r}+\hat{x}}
\end{array}\right){\rm e}^{2{\rm i}\phi},\label{Eq_S1}\\
&\left(\begin{array}{c}
b_{\bm r}\\
d_{{\bm r}+\hat{y}}
\end{array}\right)=S_2 \left(\begin{array}{c}
e_{\bm r}\\
f_{{\bm r}+\hat{y}}
\end{array}\right){\rm e}^{{\rm i}\phi},\label{Eq_S2}\\
&\left(\begin{array}{c}
e_{\bm r}\\
f_{{\bm r}+{\bm s}}
\end{array}\right)=S_3 \left(\begin{array}{c}
a_{\bm r}\\
c_{{\bm r}+{\bm s}}
\end{array}\right){\rm e}^{{\rm i}\phi}.\label{Eq_S3}
\end{align}

As reported in Ref. \onlinecite{ochiai2016floquet}, synthetic gauge fields can be implemented in this system, just by shifting relative positions of the link rings. The resulting S-matrix is expressed as   
\begin{align}
&S_1=\left(\begin{array}{cc}
\cos\beta & {\rm i}\sin\beta{\rm e}^{-{\rm i}A_1}\\
{\rm i}\sin\beta{\rm e}^{{\rm i}A_1} &  \cos\beta 
\end{array}\right),\\
&S_2=\left(\begin{array}{cc}
\cos\beta & {\rm i}\sin\beta{\rm e}^{-{\rm i}A_2}\\
{\rm i}\sin\beta{\rm e}^{{\rm i}A_2} &  \cos\beta 
\end{array}\right),\\
&S_3=\left(\begin{array}{cc}
\cos\delta & {\rm i}\sin\delta{\rm e}^{ -{\rm i}A_3}\\
{\rm i}\sin\delta{\rm e}^{{\rm i}A_3} &  \cos\delta 
\end{array}\right), \label{Eq_Smatrix_z}\\
&A_i={\bm A}\cdot{\bm a}_i,\\
&{\bm a}_1=\hat{x}, \quad  {\bm a}_2=\hat{y}, \quad {\bm a}_3={\bm s}=\frac{1}{2}(\hat{x}+\hat{y}+\hat{z}), 
\end{align}
where vector ${\bm A}$ denotes a synthetic gauge field. 
Here, we assume that the inversion symmetry holds at ${\bm A}=0$. 
The gauge symmetry under 
\begin{align}
&\alpha_{\bm r}\to \alpha_{\bm r}{\rm e}^{{\rm i}\theta_{\bm r}},\\
&A_i \to A_i+\theta_{{\bm r}+{\bm a}_i} -\theta_{\bm r},
\end{align}
is manifest.

To see how the gauge field emerges in our model, we decompose the hopping S-matrix $S_1$ into the nodal S-matrices at nodes A and C [see Fig. \ref{Fig_setup} (c)]. They are defined by 
\begin{align}
&\left(\begin{array}{c}
a_{\bm r}\\
\alpha 
\end{array}\right)=S_{\rm A} \left(\begin{array}{c}
d_{\bm r}{\rm e}^{2{\rm i}\phi}, \\
\gamma {\rm e}^{{\rm i}\phi_2}
\end{array}\right),\\
&\left(\begin{array}{c}
c_{{\bm r}+\hat{x}}\\
\gamma 
\end{array}\right)=S_{\rm C} \left(\begin{array}{c}
b_{{\bm r}+\hat{x}} {\rm e}^{2{\rm i}\phi}\\
\alpha {\rm e}^{{\rm i}\phi_1}
\end{array}\right). 
\end{align} 
By eliminating mode amplitudes $\alpha$ and $\gamma$ in the link ring, we obtain 
\begin{align}
&S_1=\left(\begin{array}{cc}
S_1^{++} & S_1^{+-} \\
S_1^{-+} & S_1^{--}
\end{array}\right),\\
&S_1^{++}=S_{\rm A}^{++} +{\rm e}^{{\rm i}(\phi_1+\phi_2)}S_{\rm A}^{+-} (1-{\rm e}^{{\rm i}(\phi_1+\phi_2)} S_{\rm C}^{--}S_{\rm A}^{--})^{-1} \nonumber\\
&\hskip50pt \times S_{\rm C}^{--}S_{\rm A}^{-+}, \\
&S_1^{+-}={\rm e}^{{\rm i}\phi_2}S_{\rm A}^{+-} (1-{\rm e}^{{\rm i}(\phi_1+\phi_2)} S_{\rm C}^{--}S_{\rm A}^{--})^{-1} S_{\rm C}^{-+},\\
&S_1^{-+}={\rm e}^{{\rm i}\phi_1}S_{\rm C}^{+-} (1-{\rm e}^{{\rm i}(\phi_1+\phi_2)} S_{\rm A}^{--}S_{\rm C}^{--})^{-1} S_{\rm A}^{-+},\\
&S_1^{--}=S_{\rm C}^{++} +{\rm e}^{ {\rm i}(\phi_1+\phi_2)}S_{\rm C}^{+-} (1-{\rm e}^{{\rm i}(\phi_1+\phi_2)} S_{\rm A}^{--}S_{\rm C}^{--})^{-1} \nonumber \\
&\hskip50pt \times S_{\rm A}^{--}S_{\rm C}^{-+}.
\end{align}
If $\phi_1+\phi_2$ is kept fixed, we obtain $A_1=(\phi_1-\phi_2)/2$. 
Since $\phi_1$ and $\phi_2$ are proportional to the corresponding arc lengths of the link ring,  nonzero $A_1$ is obtained by shifting the link rings vertically in Fig. \ref{Fig_setup} (c).   The inversion symmetry holds provided $S_{\rm A}=S_{\rm C}$ and $\phi_1=\phi_2$.

At  ${\bm A}=0$, the eigenvalue equation of the bulk modes becomes \cite{ochiai2016floquet}
\begin{align}
&U_{\bm k}\psi_{\bm k}={\rm e}^{-4{\rm i}\phi}\psi_{\bm k}, \label{Eq_Floquet} 
\end{align}                      
where ${\bm k}$ is a Bloch momentum. 
It is the diagonalization a $2\times 2$ unitary matrix $U_{\bm k}$, so that it is a Floquet form and is analytically solvable. 
 The propagation phase $\phi$ in the site rings acts  as the quasienergy of the Floquet hamiltonian $H_{\bm k}={\rm i}\log U_{\bm k}$.       
The resulting quasienergy spectrum exhibits robust Weyl points and topological or nontopological band gaps, depending on the parameters $\beta$ and $\delta$.  They correspond to the FW, FTI, and NI phases of the system, respectively.

\section{Synthetic gauge field}

\subsection{3D Hofstadter butterfly}

From now on, we impose a spatially uniform synthetic magnetic field ${\bm B}$ of fractional flux per unit area. 
For instance,  the synthetic magnetic field of flux $2\pi P_x/Q_x(=B_x)$ with coprime integers $P_x$ and $Q_x$ in the $x$ direction 
can be implemented by a Landau gauge potential of ${\bm A}= (0,-B_xz,0)$. 
Such a gauge field is obtained by shifting the link rings relevant to $S_2$ and $S_3$ systematically according to the $z$ coordinate.  
The system has the translational invariance under  $z\to z+2Q_x$, because the gauge field appears as $\exp(\pm {\rm i}A_3)=\exp(\mp {\rm i} \pi P_x/Q_x z)$
in Eq. (\ref{Eq_Smatrix_z}). 
Besides, the system has the lattice translations, $x \to x+1$ and $y \to y+1$.  
Owing to this enlarged unit cell, the Brillouin zone shrinks to the magnetic Brillouin zone (MBZ) defined by  
\begin{align}
|k_x|\le \pi, \quad |k_y|\le \pi, \quad |k_z|\le \frac{\pi}{2Q_x}, 
\end{align}
where ${\bm k}$ is the Bloch momentum. As a result, original two bands of eigenstates (at ${\bm B}=0$) split into subbands.

Under a magnetic field of fractional flux, the eigenvalue equation of the Bloch modes in the bulk is given by a Floquet-Bloch form, Eq. (\ref{Eq_Floquet}),  
with unitary matrix $U_{\bm k}$ and column vector $\psi_{\bm k}$ of finite order.          
The explicit forms of $U_{\bm k}$ and $\psi_{\bm k}$ depend on the gauge choice and are given in Appendix A.

\begin{figure}[H]
\centerline{\includegraphics[width=0.45\textwidth]{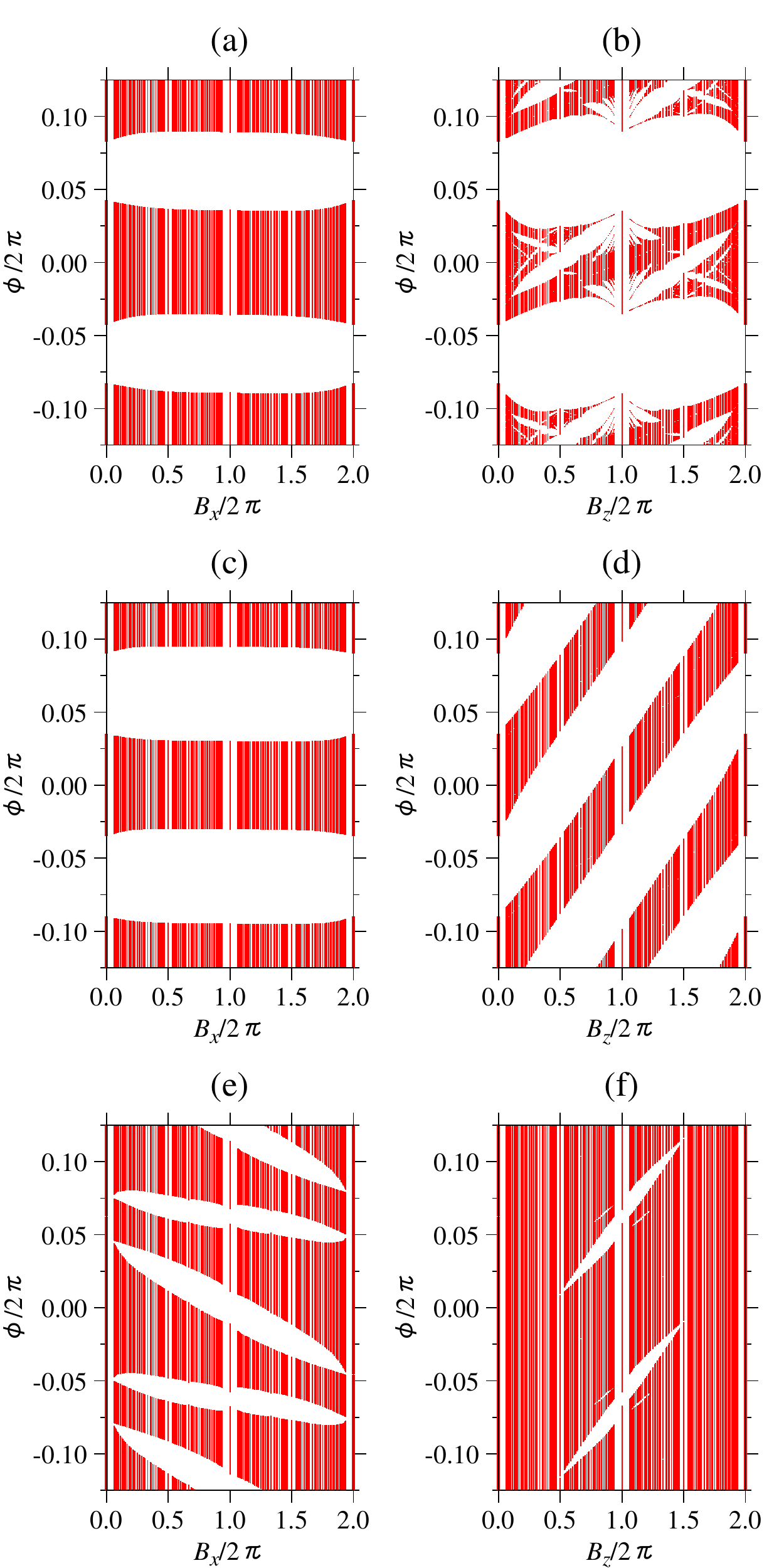}}
%\centerline{\includegraphics[width=0.65\textwidth]{fig2.pdf}}
\caption{\label{Fig_gapmap}
(Color online) Gap maps of the bulk eigenmodes as a function of magnetic flux. 
In (a) and (b), the parameters in the S-matrices are $(\beta,\delta)/2\pi=(0.43,0.47)$, being the normal-insulator phase at zero magnetic field ${\bm B}$. In (c) and (d),  $(\beta,\delta)/2\pi=(0.23,0.40)$, which are in the Floquet-topological-insulator phase at ${\bm B}=0$. In (e) and (f), $(\beta,\delta)/2\pi=(0.15,0.31)$, which are in the Floquet-Weyl phase at ${\bm B}=0$. The magnetic field is parallel to
the $x$ direction in (a), (c), and (e), and to the $z$ direction in (b), (d), and (f). Fractional fluxes of $B_\mu=2\pi P_\mu/Q_\mu$ with $Q_\mu \le 16$ are taken into account. The momentum is scanned in the entire magnetic Brillouin zone. }
\end{figure}

By scanning the magnetic flux, we obtain a gap map of the bulk eigenmodes, namely, forbidden quasienergy ranges as a function of the magnetic field.  
Figure \ref{Fig_gapmap} shows the gap map for three sets of S-matrix parameters $(\beta,\delta)$ that correspond to NI, FTI, and FW phases at ${\bm B}=0$. Two orientations  of the magnetic field, parallel to the $x$ and $z$ directions, are considered.  
%%% Fig. 2 %%%%%%%%%%%%%%%%%%%%%%%%%%%%%%%%%%%%%%%%%%%%%%%%%%%%%%%%%%%%%
%%%%%%%%%%%%%%%%%%%%%%%%%%%%%%%%%%%%%%%%%%%%%%%%%%%%%%%%%%%%%%%%%%%%

The gap map exhibits a periodicity of $4\pi$ in the flux. 
This $4\pi$ periodicity comes from the lattice structure of the network model.

It is remarkable that in Fig.~\ref{Fig_gapmap}~(b), a Hofstadter-butterfly-type spectrum is obtained. Similar spectra are obtained in 3D quantum Hall systems \cite{PhysRevB.45.13488,PhysRevLett.86.1062}. We should emphasize that our system is not an electronic system under a magnetic field, but is a bosonic system under a synthetic magnetic field. There is no need of real and strong-enough magnetic field in order to obtain such a remarkable spectrum. We just need a shifting of the link rings in the stacked 2D ring network.       
Butterfly-type spectra are not limited in Fig. \ref{Fig_gapmap} (b), where the system is in the NI phase at ${\bm B}=0$.   The other figures  also include  Hofstadter-butterfly-type ones, but they are hidden. Butterfly spectra are clearly visible in all the cases, if we restrict the Bloch momenta in a 2D plane perpendicular to the magnetic field.  
 By scanning all the  momenta in the 3D MBZ as we did in Fig. \ref{Fig_gapmap}, the butterfly spectra are overlaid and finally hidden in the figures.

In Fig. \ref{Fig_gapmap} (d), a quasienergy winding with flux is observed. Namely, the band quasienergies increase with increasing flux $B_z$, and reach the $\pi/2$ shift at $4\pi$ flux.  Since the quasienergy eigenvalue $\phi$ is periodic under $\pi/2$ shift [see Eq. (\ref{Eq_Floquet})],  they wind one period in their quasienergies. We found that the quasienergy winding is inherent in the system of the FTI phase at zero magnetic field, and in the magnetic field parallel to the $z$ direction.  If the system is in the NI or FW phase, or if the synthetic magnetic field is in the $x$ and $y$ directions, no winding is observed.

\subsection{Nontrivial topology and bulk-edge correspondence}

If the synthetic magnetic field is nonzero, we expect that the Chern number characterizes possible topological phases. 
The Chern number $C_{n\mu}$ of the $n$-th bulk band and orientation $\mu(=x,y,z)$ at the cross section of constant $k_\mu$ is defined by 
\begin{align}
&C_{n\mu}(k_\mu)=\int_{\rm MBZ} \frac{{\rm d}^2{\bm k}_\perp}{2\pi} ({\bm \nabla}_{\bm k} \times {\bm A}_{n{\bm k}})_\mu,\\
&{\bm A}_{n{\bm k}}=-{\rm i}\langle \psi_{n{\bm k}}|{\bm \nabla}_{\bm k}| \psi_{n{\bm k}}\rangle, 
\end{align}
where ${\bm k}_\perp$ is the momentum perpendicular to the $\mu$ direction, and the integral is taken over the constant $k_\mu$ plane in the MBZ. 
If the $n$-th band is fully separated from the other bands irrespective of $k_\mu$, $C_{n\mu}(k_\mu)$ does not depend on $k_\mu$ and becomes an integer. 
We refer to the integer as the Chern number of the band.

However, in many cases, the Chern number of the $n$-th band is ill-defined owing to a degeneracy with other bands at certain $k_\mu$.  Actually, the cross-sectional Chern number $C_{nx}(k_x)$ is written as 
\begin{widetext}
\begin{align}
&C_{nx}(k_x)=\int_{\rm MBZ} \frac{{\rm d}k_y{\rm d}k_z}{2\pi} \sum_{m\ne n}\frac{-{\rm i}}{({\rm e}^{-4{\rm i}\phi_n}-{\rm e}^{-4{\rm i}\phi_m})^2} \nonumber \\
&\times \left[\left(\psi_{n{\bm k}}^\dagger\frac{\partial U_{\bm k}}{\partial k_y}\psi_{m{\bm k}}\right) \left(\psi_{m{\bm k}}^\dagger\frac{\partial U_{\bm k}}{\partial k_z}\psi_{n{\bm k}}\right) - \left(\psi_{n{\bm k}}^\dagger\frac{\partial U_{\bm k}}{\partial k_z}\psi_{m{\bm k}}\right) \left(\psi_{m{\bm k}}^\dagger\frac{\partial U_{\bm k}}{\partial k_y}\psi_{n{\bm k}}\right)  \right],\nonumber \\
\label{Eq_chern_single}
\end{align} 
\end{widetext}
which has the singularity if the degeneracy occurs.

\begin{figure}
%\centerline{\includegraphics[width=0.5\textwidth]{fig3.pdf}}
\centerline{\includegraphics[width=0.3\textwidth]{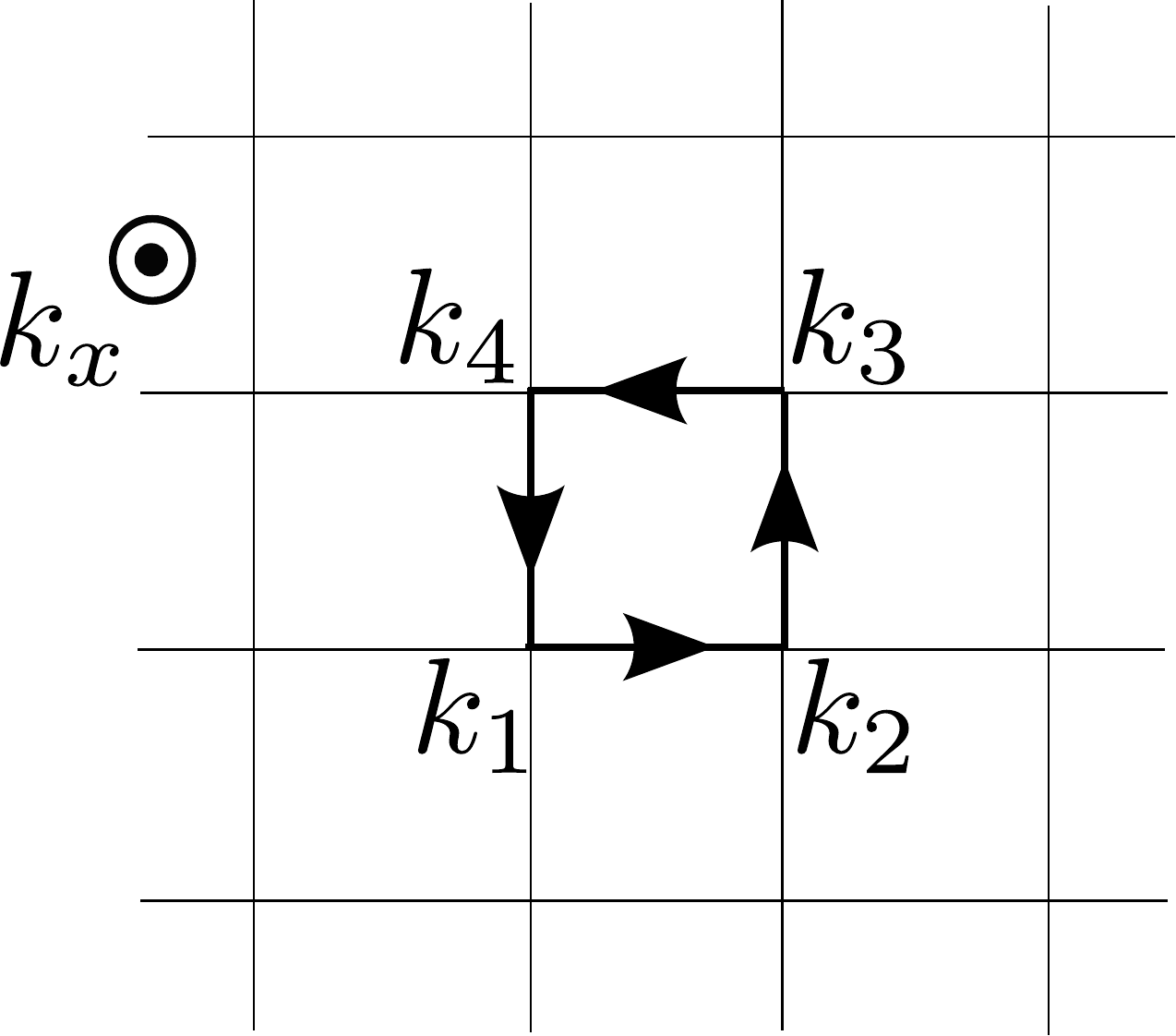}}
\caption{\label{Fig_chern_plaq}
Schematic illustration of the discretized magnetic Brillouin zone. At a cross section of constant $k_x$, the discretized Brillouin zone consists of plaquettes. As a sum over plaquettes, the bunch Chern number is evaluated as Eq. (\ref{Eq_chern_fukui}).  }
\end{figure}

To overcome this singularity, we employ the method given in Ref. \onlinecite{fukui2005cnd} to calculate the sum of the Chern numbers for a bunch of the bulk bands that may degenerate at some points in the MBZ.  
The singularity in Eq. (\ref{Eq_chern_single}) is canceled there. 
The sum of the Chern numbers (bunch Chern numbers) are evaluated as   
\begin{align}
&C_{{\rm b}x}(k_x)=\frac{1}{2\pi}\sum_{\rm plaq.}\Im[\log(D_{12}D_{23}D_{34}D_{41})],\label{Eq_chern_fukui}\\
&D_{ij}={\rm det}(\langle \psi_{m{\bm k}_i}|\psi_{n{{\bm k}_j}}\rangle ),  
\end{align}
for the $x$ direction.  
Here, the MBZ is discretized with a uniform mesh as shown in Fig. \ref{Fig_chern_plaq}, and the summation is taken over all the plaquettes on the cross section of fixed $k_x$. The indices $m$ and $n$ are the band indices and run over the bunch of the bulk bands that may degenerate with each other. The number of the bulk bands in the bunch is denoted as $N_{\rm b}$. If the bunch of the bulk bands is separated in quasienergy from the other bands, Eq. (\ref{Eq_chern_fukui}) becomes an integer irrespective of $k_\mu$.  
%%% Fig.3 %%%%%%%%%%%%%%%%%%%%%%%%%%%%%%%%%%%%%%%%%%%%%%%%
%%%%%%%%%%%%%%%%%%%%%%%%%%%%%%%%%%%%%%%%%%%%%%%%%%%%%%%%%%%%%

Besides, the system is also characterized by the winding number given  by
\begin{align}
n_{\mu\nu}^{({\rm l})}=\int_{\rm MBZ} \frac{{\rm d}k_\nu}{2\pi}\frac{\partial \Im[{\rm log}({\rm det}R_\mu^{({\rm l})})]}{\partial k_\nu},  
\end{align}
where $R_\mu^{({\rm l})}$ is the reflection matrix of the semi-infinite system whose surface is  normal to the $\mu$ direction \cite{ochiai2016floquet}. The reflection matrix is obtained from the slab S-matrix of the system as follows. Suppose that  
 the slab has finite-thickness in the direction parallel to $\hat{\mu}$ and infinite extent in the direction normal to $\hat{\mu}$. We denote the slab S-matrix as $S_{\mu;N}$ for $N$-layer thick slab. In a pseudo gap of the bulk eigenmodes, the transmission (diagonal) block of the S-matrix vanishes as $N\to\infty$. Therefore, the S-matrix is then written as 
\begin{align}
S_{\mu;N}\to \left(\begin{array}{cc}
0&R_\mu^{({\rm u})} \\
R_\mu^{({\rm l})} & 0 
\end{array}\right), 
\end{align}
with the reflection matrices $R_\mu$ of the upper [superscript $({\rm u})$] and lower [superscript $({\rm l})$] slab surfaces. 
An explicit construction of the slab S-matrix is given in Appendix B.

The winding number is defined in a pseudogap of the bulk bands and is directly related to the number of chiral surface states  in the pseudogap.  Such a topological characterization via winding number relevant to S-matrix is universal and employed in various systems \cite{PhysRevB.58.R10135,braunlich2010equivalence,PhysRevB.85.165409}.

\begin{figure}
\centerline{\includegraphics[width=0.5\textwidth]{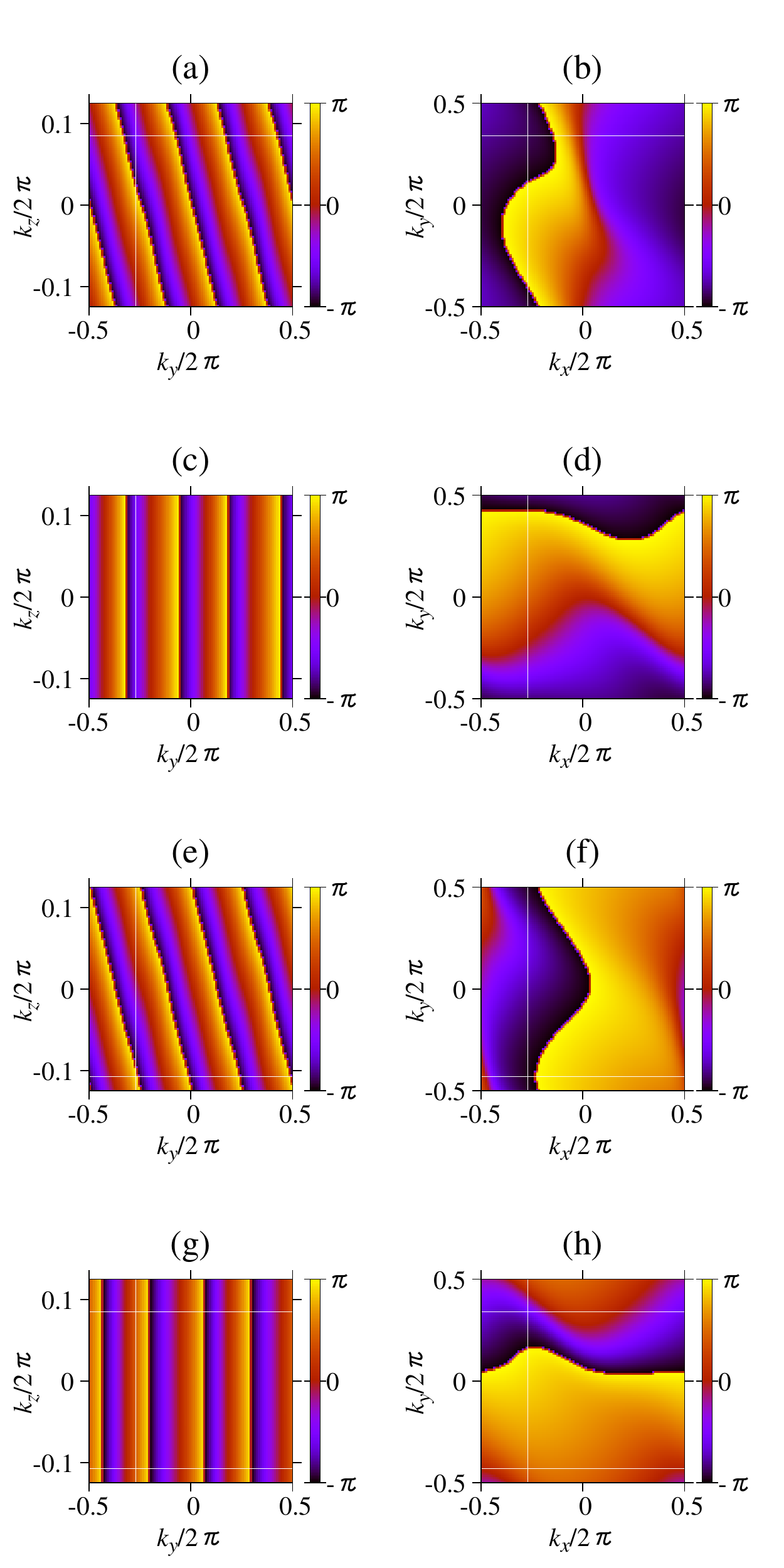}}
%\centerline{\includegraphics[width=0.65\textwidth]{fig4.pdf}}
\caption{\label{Fig_winding} (Color online)  Winding features of $\Im[\log({\rm det}R_\mu^{({\rm l})})]$ for $\pi$-flux magnetic field in the $x$ direction. 
The gauge choice is ${\bm A}=(0,-B_xz,0)$, and the S-matrix parameters are taken to be $(\beta,\delta)/2\pi=(0.15,0.31)$, which are in the Floquet-Weyl phase at ${\bm B}=0$. 
The boundary surface is normal to the $x$ direction in (a), (c), (e), and (g), and to the $z$ direction in (b), (d), (f), and (h). The quasienergy values are $\phi/2\pi=0.07$ [(a) and (b)], 0.025 [(c) and (d)], -0.055  [(e) and (f)], and -0.1  [(g) and (h)], which are inside the band gaps of Fig. \ref{Fig_gapmap} (e).  }
\end{figure}

Figure \ref{Fig_winding} shows $\Im[{\rm log}({\rm det}R_\mu^{({\rm l})})]$ as a function of momentum at some gap quasienergies of Fig. \ref{Fig_gapmap} (e).   
%%% Fig.4 %%%%%%%%%%%%%%%%%%%%%%%%%%%5
%%%%%%%%%%%%%%%%%%%%%%%%%%%%%%%%%%%%%%% 
The figure clearly indicates winding features of the reflection matrices. 
Namely, $\Im[{\log}({\rm det}R_\mu^{({\rm l})})]$ changes from $-\pi$ to $\pi$, zero, one or four times in the figure, as the momentum traverses the MBZ. 
The winding number depends on the gap concerned and the surface orientation.

Nonzero winding number results in the emergence of gapless surface states. 
The dispersion relation of the surface states are given by 
\begin{align}
{\rm det}(1-R_\mu^{({\rm l})}T_\mu^{({\rm l})})=0,  
\end{align} 
for the lower surface of the slab, 
where $T_\mu^{({\rm l})}$ represents the boundary condition at the surface. 
Both $R_\mu^{({\rm l})}$ and $T_\mu^{({\rm l})}$ are unitary, and their expressions are given in Appendix B.

Figure \ref{Fig_SS} shows the dispersion relation of the surface states for the $\pi$-flux magnetic field in the $x$ direction and for the surface normal to the $x$ or $z$ direction.    
%%%% Fig. 5 %%%%%%%%%%%%%%%%%%%%%%%%%%%%%%%%%%%%%%%%%%%%%%
\begin{figure}
\centerline{\includegraphics[width=0.5\textwidth]{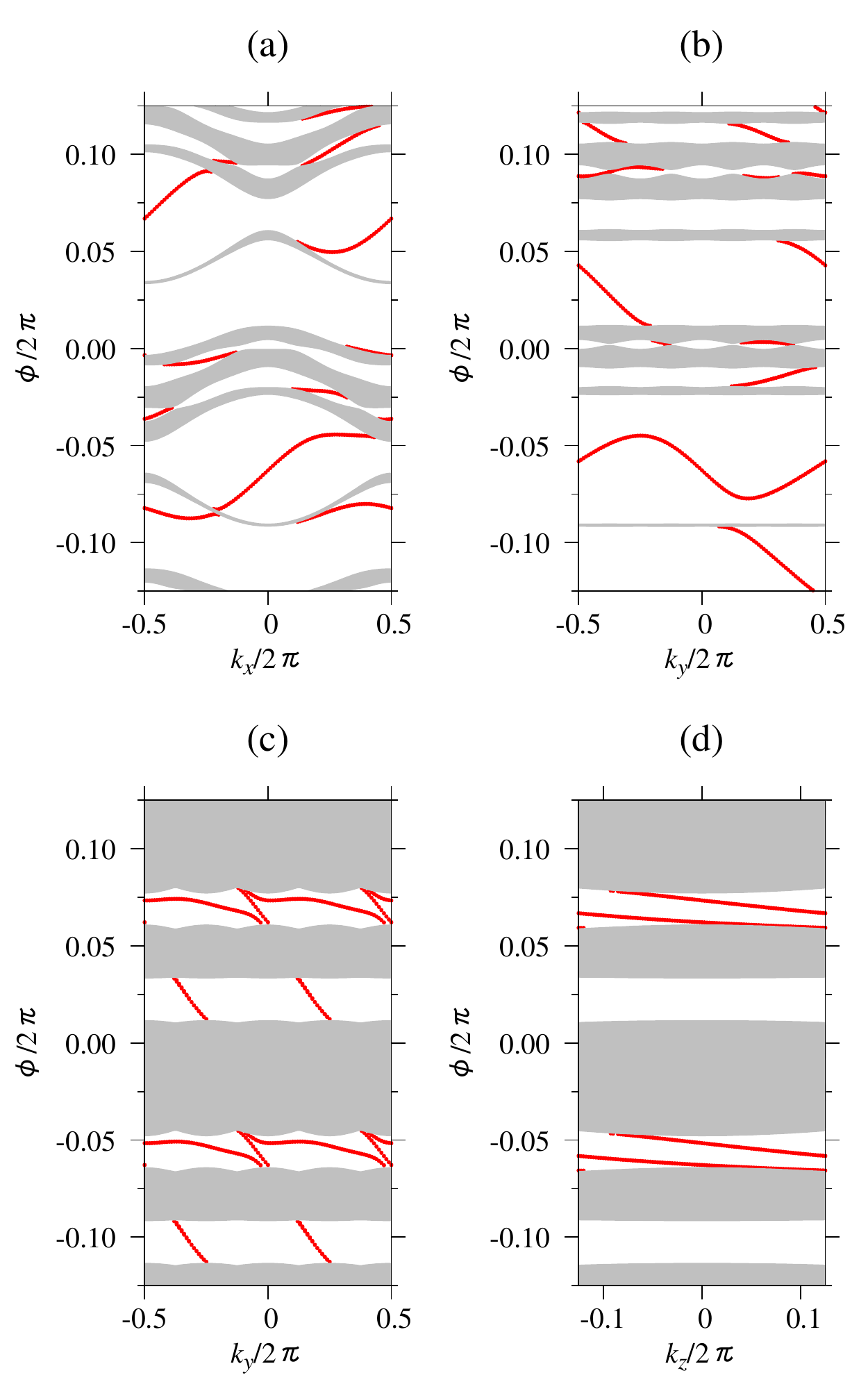}}
%\centerline{\includegraphics[width=0.8\textwidth]{fig5.pdf}}
\caption{\label{Fig_SS} (Color online) Band structure of the lower surface states under the $\pi$-flux magnetic field in the $x$ direction. The surface is normal to the $z$ direction in (a) and (b), and to the $x$ direction in (c) and (d).  Solely the dispersion relation on $k_y=0$ (a), $k_x=0$ (b), $k_z=0$ (c), and $k_y=0$ (d) is assumed. 
The gauge choice is ${\bm A}=(0,-B_xz,0)$, and the S-matrix parameters are taken to be $(\beta,\delta)/2\pi=(0.15,0.31)$.  The additional phases $\varphi_\alpha$ ($\alpha=$A, B, C, D, E, F) at the boundary surfaces, which appear in the boundary S-matrices $T_\mu^{({\rm l})}$, are taken to be zero.  
The shaded region is the projection of the bulk band structure. In (c), the dispersion curves around  $\phi/2\pi=0.025$ and -0.10 are almost doubly degenerate, so that there are four surface states. }
\end{figure}
%%%%%%%%%%%%%%%%%%%%%%%%%%%%%%%%%%%%%%%%%%%%%%%%%%%%%%
We can find several gapless surface states in the figure. The number of the gapless surface states coincides to the winding number in the gap concerned. 
For instance, around $\phi/2\pi=0.07$ in Fig. \ref{Fig_SS} (c), we have four surface-state dispersion curves (taking account of the periodicity in the MBZ) for the $x$-oriented surface in the $k_y$ direction. This fact corresponds to $n_{xy}^{({\rm l})}=4$ indicated in Fig. \ref{Fig_winding} (a) and Table I. 
The sign of the winding number corresponds to the chirality (left or right-going) of the surface states. 

We also note that the gapless surface states emerge in the $z$-oriented surface. Such a surface state is not allowed if the synthetic magnetic field is zero \cite{ochiai2016floquet}. 
At ${\bm B}=0$, the system is like a weak topological insulator obtained by stacking 2D topological-insulator layers, so that the gapless surface states are limited in the planes parallel to the $z$ direction. However, the synthetic magnetic field can create the gapless surface states in the plane perpendicular to $z$.

The winding numbers evaluated from Fig. \ref{Fig_winding}, and the bunch Chern numbers are shown in Table \ref{Table_topoinv}, for the $\pi$-flux magnetic field in the $x$ direction. 
%%% Table I %%%%%%%%%%%%%%%%%%%%%%%%%%%%%%%%%%%%%%%%%%%%%%%%%%%%%%%%5
 \begin{table}
 \caption{\label{Table_topoinv} Chern numbers and winding numbers at $\pi$-flux magnetic field in the $x$ direction. The following parameters are employed: $(\beta,\delta)/2\pi=(0.15,0.31)$. }
\begin{center}
%\begin{ruledtabular}
\begin{tabular}{|c|c|c|c|c|c|c|c|c|}\hline
$\phi/2\pi$ & $N_{\rm b}$ & $C_{{\rm b}x}$ & $C_{{\rm b}y}$ & $C_{{\rm b}z}$ & $n_{xy}^{({\rm l})}$ & $n_{xz}^{({\rm l})}$ &  $n_{zx}^{({\rm l})}$ & $n_{zy}^{({\rm l})}$ \\ \hline
(0.0766,0.1250) & & & & & \multicolumn{4}{c|}{} \\
(-0.1250,-0.1132)&6 & -1 & 1 & 0 & \multicolumn{4}{c|}{} \\ \hline
{\rm gap}& \multicolumn{4}{c|}{} & 4 & 1 & -1 & 0 \\ \hline
(0.0334,0.0611)&2 &  1 & -1& 0 & \multicolumn{4}{c|}{} \\ \hline
{\rm gap}& \multicolumn{4}{c|}{} & 4 & 0 & 0 & 1 \\ \hline
(-0.0483,0.0118)&6 & -1 & 1 & 0 & \multicolumn{4}{c|}{}  \\ \hline
{\rm gap}& \multicolumn{4}{c|}{} & 4 & 1 & -1 & 0 \\ \hline
(-0.0916,-0.0639)&2 &  1 & -1& 0 & \multicolumn{4}{c|}{} \\ \hline
{\rm gap}& \multicolumn{4}{c|}{} & 4 & 0 & 0 & 1 \\ \hline 
\end{tabular}
%\end{ruledtabular}
\end{center}
\end{table}
%%%%%%%%%%%%%%%%%%%%%%%%%%%%%%%%%%%%%%%%%%%%%%555

We found that the following relation of the bulk-edge correspondence holds in all the cases examined so far: 
\begin{align}
n_{\mu\nu}^{({\rm l})+} - n_{\mu\nu}^{({\rm l})-} = \epsilon_{\mu\nu\rho}C_{{\rm b}\rho}. 
\end{align}
where $n_{\mu\nu}^{({\rm l})\pm}$ is the winding number of the reflection matrix of semi-infinite thickness in the upper (superscript:+) and lower (-) gaps of the bulk bands whose bunch Chern number is ${\bm C}_{\rm b}$.  A similar relation of the bulk-edge correspondence in 2D Floquet-Bloch systems was derived \cite{PhysRevX.3.031005}. We also found $n_{\mu\nu}^{({\rm u})}=-n_{\mu\nu}^{({\rm l})}$.

We should note that a vanishing Chern number is fully consistent with nonzero winding numbers, or in other words, the presence of the chiral surface states. 
Namely, if $n_{\mu\nu}^+ = n_{\mu\nu}^-\ne0$ for all the band, then  the number of the chiral surface states is the same among the all gaps.

\section{Pseudospin-orbit interaction}

\subsection{Model}
So far, we have assumed the case that the counter-clockwise mode is preserved in the site rings. In our system, we also have the clockwise mode. Therefore, we have assumed that they do not mix with each other. This assumption can be justified in the Lieb lattice structure together with the directional coupling at the nodes. 
  
However, it is sometimes the case that the directional coupling is not preserved exactly. In this case, we have to consider the mixing among the pseudospin degrees of freedom.   

The S-matrices that include the both pseudospin degrees are given by 
\begin{align}
&\left(\begin{array}{l}
a_{\bm r}  \\
a'_{\bm r}  \\
c_{{\bm r}+\hat{x}}  \\
c'_{{\bm r}+\hat{x}}  
      \end{array}\right)=S_1
\left(\begin{array}{l}
d_{\bm r}{\rm e}^{2{\rm i}\phi}  \\
e'_{\bm r}{\rm e}^{{\rm i}\phi}  \\
b_{{\bm r}+\hat{x}}{\rm e}^{2{\rm i}\phi}  \\
f'_{{\bm r}+\hat{x}}{\rm e}^{{\rm i}\phi} 
      \end{array}\right),\label{Eq_S1_PSOI}\\
&\left(\begin{array}{l}
b_{\bm r}  \\
b'_{\bm r}  \\
d_{{\bm r}+\hat{y}}  \\
d'_{{\bm r}+\hat{y}}  
      \end{array}\right)=S_2
\left(\begin{array}{l}
e_{\bm r}{\rm e}^{{\rm i}\phi}  \\
c'_{\bm r}{\rm e}^{2{\rm i}\phi}  \\
f_{{\bm r}+\hat{y}}{\rm e}^{{\rm i}\phi}  \\
a'_{{\bm r}+\hat{y}}{\rm e}^{  2{\rm i}\phi} 
      \end{array}\right),\label{Eq_S2_PSOI}\\
&\left(\begin{array}{l}
e_{\bm r}  \\
e'_{\bm r}  \\
f_{{\bm r}+{\bm s}}  \\
f'_{{\bm r}+{\bm s}}  
      \end{array}\right)=S_3
\left(\begin{array}{l}
a_{\bm r}{\rm e}^{{\rm i}\phi}  \\
b'_{\bm r}{\rm e}^{{\rm i}\phi}  \\
c_{{\bm r}+{\bm s}}{\rm e}^{{\rm i}\phi}  \\
d'_{{\bm r}+{\bm s}}{\rm e}^{{\rm i}\phi} 
      \end{array}\right),\label{Eq_S3_PSOI}
\end{align}
where $\alpha_{\bm r}$ and  $\alpha'_{\bm r}$ are counter-clockwise and clockwise mode amplitudes, respectively, as shown in Fig. \ref{Fig_setup}.

The eigenvalue equation of the bulk modes becomes 
\begin{align}
&U_{\bm k}\psi_{\bm k}={\rm e}^{-2{\rm i}\phi}\psi_{\bm k},\\ 
&U_{\bm k}=\left(\begin{array}{llll}
\tilde{S}_1^{+-}\tilde{S}_3^{-+}& \tilde{S}_1^{++}& 0 & \tilde{S}_1^{+-}\tilde{S}_3^{--}\\
\tilde{S}_2^{++}\tilde{S}_3^{++}& 0 & \tilde{S}_2^{+-}& \tilde{S}_2^{++}\tilde{S}_3^{+-}\\
\tilde{S}_1^{--}\tilde{S}_3^{-+}& \tilde{S}_1^{-+}& 0 & \tilde{S}_1^{--}\tilde{S}_3^{--}\\
\tilde{S}_2^{-+}\tilde{S}_3^{++}& 0 & \tilde{S}_2^{--}& \tilde{S}_2^{-+}\tilde{S}_3^{+-}
\end{array}\right),\\
&\tilde{S}_1=\left(\begin{array}{cccc}
1& 0& 0& 0\\
0&0&{\rm e}^{-{\rm i}k_x}&0\\
0&1&0&0\\
0&0&0&{\rm e}^{-{\rm i}k_x}
\end{array}\right)S_1
\left(\begin{array}{cccc}
0& 1& 0& 0\\
0&0&1&0\\
{\rm e}^{{\rm i}k_x}&0&0&0\\
0&0&0&{\rm e}^{{\rm i}k_x}
\end{array}\right),\label{Eq_S1t_PSOI}\\
&\tilde{S}_2=\left(\begin{array}{cccc}
1& 0& 0& 0\\
0&0&{\rm e}^{-{\rm i}k_y}&0\\
0&1&0&0\\
0&0&0&{\rm e}^{-{\rm i}k_y}
\end{array}\right)S_2
\left(\begin{array}{cccc}
1& 0& 0& 0\\
0&0&0&1\\
0&{\rm e}^{{\rm i}k_y}&0&0\\
0&0&{\rm e}^{{\rm i}k_y}&0
\end{array}\right),\label{Eq_S2t_PSOI}\\
&\tilde{S}_3=\left(\begin{array}{cccc}
1& 0& 0& 0\\
0&0&{\rm e}^{-{\rm i}k_3}&0\\
0&1&0&0\\
0&0&0&{\rm e}^{-{\rm i}k_3}
\end{array}\right)S_3
\left(\begin{array}{cccc}
1& 0& 0& 0\\
0&0&1&0\\
0&{\rm e}^{{\rm i}k_3}&0&0\\
0&0&0&{\rm e}^{{\rm i}k_3}
\end{array}\right),\label{Eq_S3t_PSOI}\\
&\tilde{S}_j=\left(\begin{array}{cc}
\tilde{S}_j^{++}& \tilde{S}_j^{+-}\\
\tilde{S}_j^{-+}& \tilde{S}_j^{--}
\end{array}\right),\\
&\psi_{\bm k}=(a_{\bm r},c_{\bm r},b_{\bm r},d_{\bm r},a'_{\bm r},c'_{\bm r},b'_{\bm r},d'_{\bm r})^{\rm t},\\
&k_3={\bm k}\cdot{\bm s}. 
\end{align}

Compared with the system without the PSOI, the system with the PSOI lacks the invariance under $\phi\to\phi + \pi/2$. The periodicity becomes doubled: $\phi\to\phi + \pi$.

Owing to the coexistence of the both clockwise and counterclockwise modes, the TRS becomes clarified. If the pseudospin degrees are decoupled, the TRS seems to be broken in each pseudospin sector.  
The TRS in the S-matrices is expressed  as 
\begin{align}
S_j^{\rm t}=\left(\begin{array}{cc}
\sigma_1&0 \\
0&\sigma_1
\end{array}\right)S_j
\left(\begin{array}{cc}
\sigma_1&0 \\
0&\sigma_1
\end{array}\right), \quad (j=1,2,3), \label{Eq_Smat_TRS}
\end{align}
where $\sigma_j$ is the Pauli matrix.   
In terms of the eigenvalue equation, the TRS is represented by
\begin{align}
&{\cal T}^{-1}U_{-{\bm k}}{\cal T}=U_{\bm k}^{-1},\\ 
&{\cal T}=\left(\begin{array}{cccc}
(\tilde{S}_3^{-+})^* & 0 & 0 & (\tilde{S}_3^{--})^* \\
0 & 0 & \sigma_1 & 0\\
0 & \sigma_1 & 0 & 0\\
(\tilde{S}_3^{++})^* & 0 & 0 & (\tilde{S}_3^{+-})^* 
\end{array}\right){\cal K},
\end{align} 
where ${\cal K}$ is the complex conjugation operator. 
The TRS is bosonic, satisfying ${\cal T}^2=1$.

Besides, the S-matrices hold the inversion symmetry with respect to the center of the link ring and the mirror symmetry with respect to the plane bisecting the link ring horizontally,  provided that the link ring is at the symmetric position [see Fig. \ref{Fig_setup} (c)]. 
The inversion and mirror symmetries are expressed as    
\begin{align}
&S_j=\left(\begin{array}{cc}
0&\hat{1} \\
\hat{1}&0
\end{array}\right)S_j
\left(\begin{array}{cc}
0&\hat{1} \\
\hat{1}&0
\end{array}\right),\label{Eq_Smat_SIS}\\
& S_j=\left(\begin{array}{cc}
\sigma_1&0 \\
0&\sigma_1
\end{array}\right)S_j
\left(\begin{array}{cc}
\sigma_1&0 \\
0&\sigma_1
\end{array}\right),
\end{align} 
respectively. 
Here, $\hat{1}$ is the $2\times 2$ unit matrix. 
In terms of the eigenvalue equation, these symmetries result in the SIS  ${\cal S}$ and parity ${\cal P}$ with respect to the plane $x=y$. They are represented by 
\begin{align}
&{\cal S}^{-1}U_{-{\bm k}}{\cal S}=U_{\bm k},\\ 
&{\cal S}={\rm Bdiag}(\sigma_1,\sigma_1,\sigma_1,\sigma_1),\\
&{\cal P}^{-1}U_{k_y,k_x,k_z}{\cal P}=U_{\bm k},\\ 
&{\cal P}=\left(\begin{array}{cccc}
0 & 0 & 0 & \hat{1}\\
0 & 0 & \hat{1} & 0\\
0 & \hat{1} & 0 & 0\\
\hat{1} & 0 & 0 & 0
\end{array}\right), 
\end{align}
where Bdiag stands for block diagonal.

The allowed form of the S-matrices under these symmetries is given by 
\begin{align}
&S_j={\rm e}^{{\rm i}H_j},\\ 
&H_j=\left(\begin{array}{cccc}
\alpha_j & \beta_j & \gamma_j & \delta_j   \\
\beta_j  & \alpha_j & \delta_j & \gamma_j  \\
\gamma_j & \delta_j & \alpha_j & \beta_j \\
\delta_j & \gamma_j & \beta_j & \alpha_j
\end{array}\right),
\end{align} 
with real $\alpha_j,\beta_j,\gamma_j$, and  $\delta_j$. 
The parameter $\alpha_j$ gives the overall phase to $S_j$.
The parameters  $\beta_j$ and $\delta_j$ describe the PSOI.  The parameter $\gamma_j$ represents the interaction among the rings with the same pseudospin.

\subsection{Destruction of the FTI phase}

As a consequence of the PSOI, the FTI phase of the system without the PSOI is generally destroyed. This can be understood as follows.    
At zero PSOI, we have a decoupled pair of the clockwise and counter-clockwise modes. They are related with each other by the time-reversal transformation, thus ${\bm k}$ to $-{\bm k}$ of the Bloch momentum. This is also true for the chiral surface states in the FTI phase. The dispersion curves of the chiral surface states cross at the time-reversal-invariant momenta, ${\bm k}_\|=(0,0),(\pi,0),(0,\pi)$, and $(\pi,\pi)$, where ${\bm k}_\|$ is the momentum in the surface Brillouin zone.  We should recall that the TRS is bosonic, satisfying ${\cal T}^2=1$. This bosonic nature prohibits the Kramers degeneracy at the momenta, opening a gap there. In this way, the surface states are no longer gapless, and the FTI phase is lost  by the PSOI.

\begin{figure}
\centerline{\includegraphics[width=0.3\textwidth]{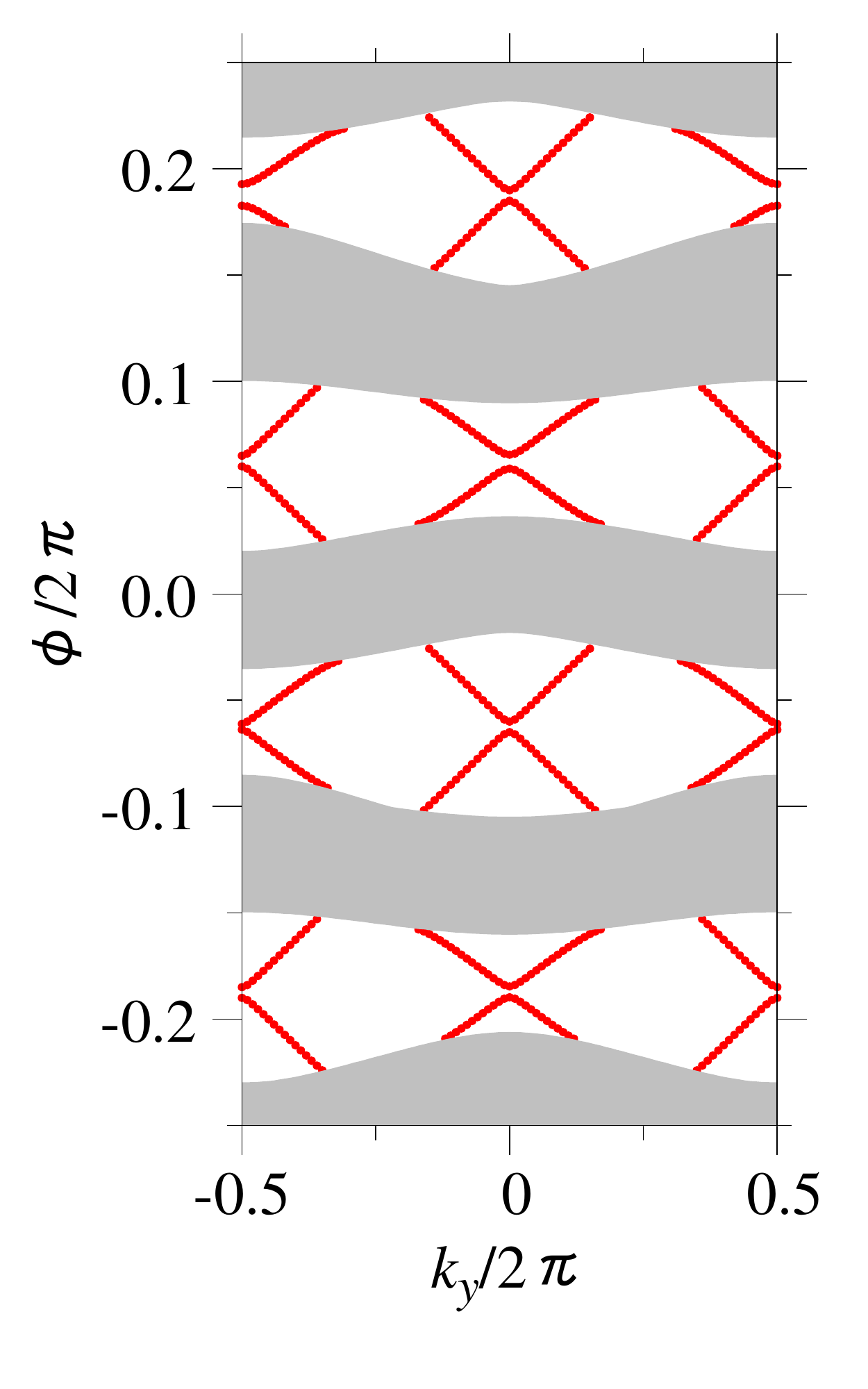}}
%\centerline{\includegraphics[width=0.6\textwidth]{fig6.pdf}}
\caption{\label{Fig_fti_soi} (Color online) The band structure of the lower surface states for the system with the pseudospin-orbit interaction. The shaded region is the projection of the bulk eigenmodes.  The surface is normal to the $x$ direction and $k_z=0$ is assumed. The additional phases $\varphi_C$ and $\varphi_F$ at the boundary surface are taken to be zero. 
The following parameters of the S-matrices are employed: $(\alpha_1,\beta_1,\gamma_1,\delta_1)/2\pi=(\alpha_2,\beta_2,\gamma_2,\delta_2)/2\pi=(0,0.01,0.23,0.01)$, $(\alpha_3,\beta_3,\gamma_3,\delta_3)/2\pi=(0,0.01,0.40,0.01)$. }
\end{figure}

Figure \ref{Fig_fti_soi} shows the dispersion curve of the surface states for the surface normal to the $x$ direction.  
Here, we add small $\beta_j$ and $\delta_j$ as a perturbation of the PSOI.  Without these terms, the system is in the FTI phase with gapless surface states.  
%%% FIG. 6 %%%%%%%%%%%%%%%%%%%%%%%%%%%%%%%%%%%%%%%%%%%%%%%
%%%%%%%%%%%%%%%%%%%%%%%%%%%%%%%%%%%%%%%%%%%%%%%%%%%%%%
We can see the gaps of the surface states at the time-reversal invariant momenta. The surface states are thus no longer gapless, and the FTI phase is destroyed.   
Actually, we can confirm that the winding numbers in the bulk band gaps are zero, so that the system becomes nontopological.

The possibility of having (Floquet) topological-crystalline-insulator phases \cite{fu2011topological,hsieh2012topological} seems to be ruled out, because the only symmetry of the surface is the parity with respect to the plane of $x=y$ for the surface normal to the $z$ direction. This parity symmetry is bosonic, satisfying ${\cal P}^2=1$. Therefore, the argument of the topological crystalline insulator by a mirror symmetry \cite{hsieh2012topological} is not applicable. For the surface normal to the $x$ direction, there is no spatial symmetry other than the lattice translation.

\subsection{Weyl points and Fermi arc}

It is well known that the Weyl point cannot emerge in the systems having the both TRS and SIS. We need to break either one of the two symmetries to preserve the FW phase. The TRS is difficult to break in optical systems, because the breaking is usually through the magneto-optical effect which is very small in a wide frequency range. 
Therefore, we are forced to break the SIS.

One way to break the SIS in our system is to shift the relative position of the link rings as we did in the implementation of the synthetic gauge field. The shift must be uniform to preserve the Bloch periodicity. 
In this case, the S-matrices are written as 
\begin{align}
&S_j={\rm e}^{{\rm i}H_j},\label{Eq_S_shift1}\\ 
&H_j=\left(\begin{array}{cccc}
\alpha_j & \beta_j & \gamma_j & \delta_j   \\
\beta_j^* & \alpha_j & \delta_j & \gamma_j^*\\
\gamma_j^* & \delta_j & \alpha_j & \beta_j^*\\
\delta_j & \gamma_j & \beta_j & \alpha_j
\end{array}\right),\label{Eq_S_shift2}
\end{align}
with real $\alpha_j$ and $\delta_j$, and complex $\beta_j$ and $\gamma_j$. 
If we put $\alpha_j=\beta_j=\delta_j=0$, we have 
\begin{widetext}
\begin{align}
&S_j=\left(\begin{array}{cccc}
\cos|\gamma_j| & 0 & {\rm i}\sin|\gamma_j|{\rm e}^{{\rm i}{\rm arg}\gamma_j}& 0\\
0 & \cos|\gamma_j| & 0 & {\rm i}\sin|\gamma_j|{\rm e}^{-{\rm i}{\rm arg}\gamma_j} \\
{\rm i}\sin|\gamma_j|{\rm e}^{  -{\rm i}{\rm arg}\gamma_j} & 0 & \cos|\gamma_j| & 0 \\
0 & {\rm i} \sin|\gamma_j|{\rm e}^{ {\rm i} {\rm arg}\gamma_j} & 0 &  \cos|\gamma_j|
	  \end{array}\right). 
\end{align}
\end{widetext}
This form is equivalent to the S-matrices without the PSOI but with the synthetic gauge field of $A_j=\pm {\rm arg}\gamma_j$. The plus (minus) sign corresponds to the gauge field for the clockwise (counter-clockwise) mode.

Since the PSOI is expected to be small in the Lieb lattice structure, it is natural to consider the above form as the zeroth order approximation. Then, we introduce small $\beta_j$ and $\delta_j$ as a perturbation of the PSOI.

\begin{figure}
\centerline{\includegraphics[width=0.3\textwidth]{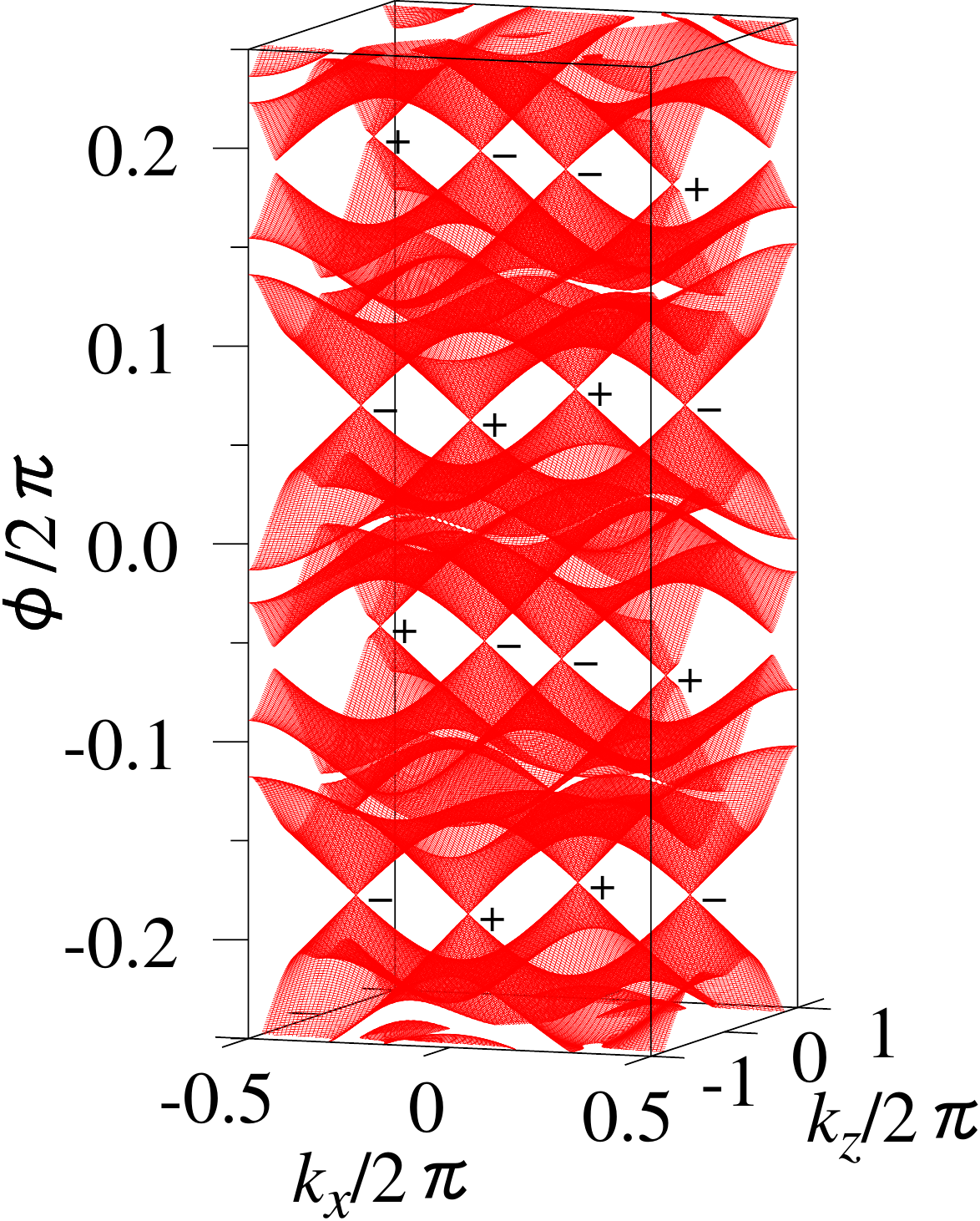}}
%\centerline{\includegraphics[width=0.8\textwidth]{fig7.pdf}}
\caption{\label{Fig_soi_weyl} (Color online) The bulk band structure including multiple Weyl points in the system with the pseudospin-orbit interaction. Solely, the plane of $k_x=k_y$ is considered. The $\pm$ sign of the Weyl points represents chirality. 
The following parameters are employed: $(\alpha_1,\beta_1,\gamma_1,\delta_1)/2\pi=(\alpha_2,\beta_2,\gamma_2,\delta_2)/2\pi=(0,0.02,0.15\exp(-{\rm i}\pi/3),0.02)$,  $(\alpha_3,\beta_3,\gamma_3,\delta_3)/2\pi=(0,0.02,0.31\exp(-{\rm i}\pi/7),0.02)$. }
\end{figure}

Figure \ref{Fig_soi_weyl} shows the typical quasienergy band structure with multiple  Weyl points in the system with the PSOI.   
%%% FIG. 7 %%%%%%%%%%%%%%%%%%%%%%%%%%%%%%%%%%%%%%%%%%%%%%%
%%%%%%%%%%%%%%%%%%%%%%%%%%%%%%%%%%%%%%%%%%%%%%%%%%%%%%
The band touching with linear dispersion is clearly observed around $\phi/2\pi=\pm 1/16, \pm 3/16$, which correspond to the exact Weyl-point quasienergy values in the system without the PSOI \cite{ochiai2016floquet}.  The Weyl points emerge on the $k_x=k_y$ plane, being in a pair of ${\bm k}_c$ and $-{\bm k}_c$ at the same $\phi_c$. 
The latter property comes from the TRS of the system. 
Around $\phi/2\pi=1/16$, for instance, we have two pairs of the Weyl points. The two pairs have slightly different $\phi_c$, owing to the small PSOI.

\begin{figure}
\centerline{\includegraphics[width=0.45\textwidth]{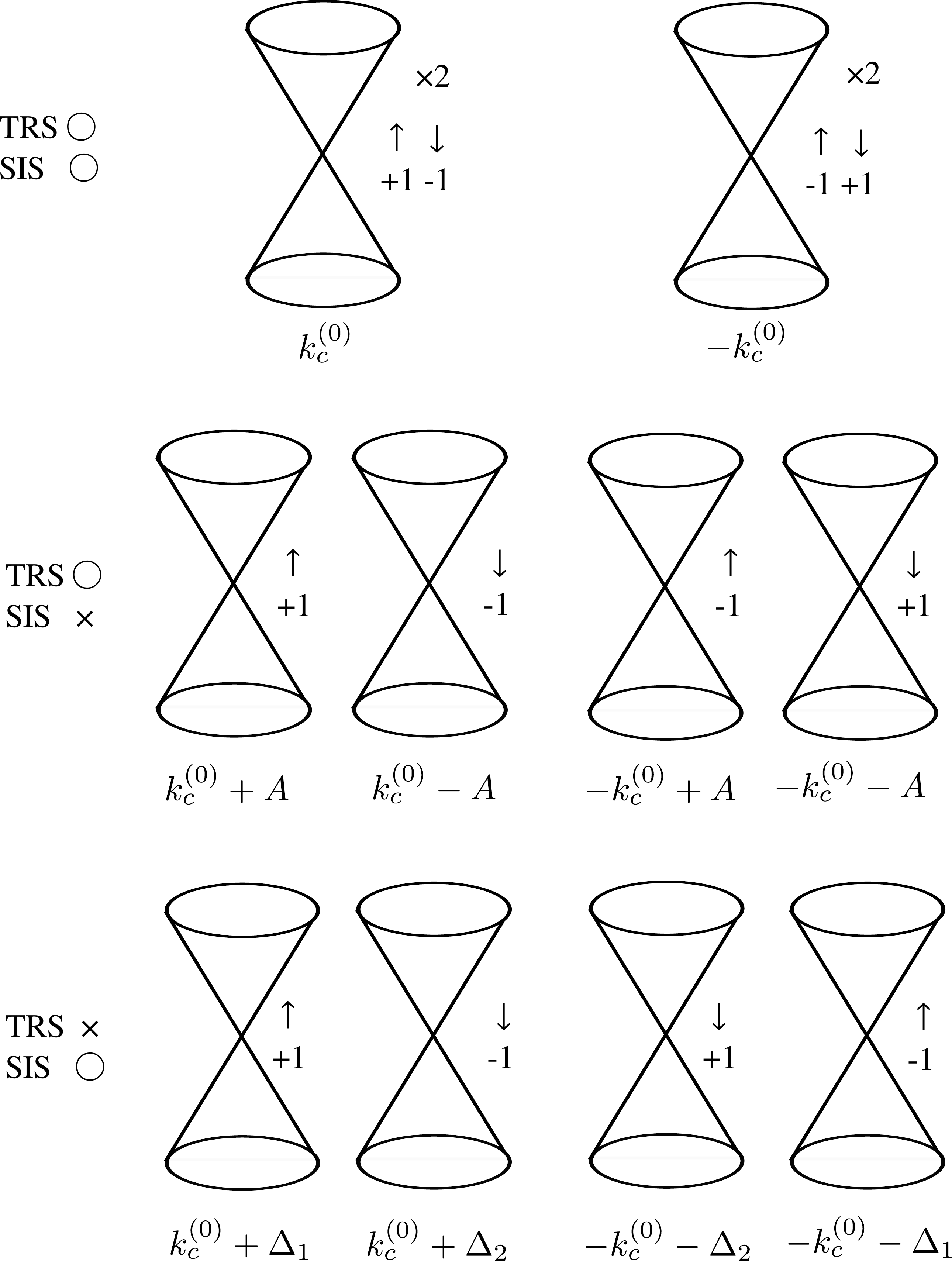}}
%\centerline{\includegraphics[width=0.8\textwidth]{fig8.pdf}}
\caption{\label{Fig_weyl_schematic}
Schematic illustration of the Weyl-point formation in the system without the PSOI. Arrow represents the pseudospin and the number below  refers to the chirality of the Weyl point. If the system has both the TRS and SIS, each pseudospin sector has a pair of the Weyl points with opposite chiralities at ${\bm k}={\bm k}_c^{(0)}$ and $-{\bm k}_c^{(0)}$. Each Weyl cone is doubly degenerate with respect to the pseudospin. By breaking the SIS via constant (and pseudospin-dependent) gauge field ${\bm A}$, the degeneracy is lifted, and the Weyl points move by $\pm {\bm A}$ in momentum space. Note that the time-reversal partners [e.g., those at $\pm ({\bm k}_c^{(0)}+{\bm A})$]  have the same chirality. On the other hand, by breaking the TRS, the degeneracy is lifted in a different way. Note that the space-inversion partners [e.g., those at $\pm ({\bm k}_c^{(0)}+{\bm \Delta}_1)$] have the opposite chiralities.} 
\end{figure}

To understand the above properties, we consider the system without the PSOI, so that the pseudospin degrees are decoupled. A schematic illustration of the Weyl-point formation is given in Fig. \ref{Fig_weyl_schematic}. 
%%% FIG. 8 %%%%%%%%%%%%%%%%%%%%%%%%%%%%%%%%%%%%%%%%%%%%%%%
%%%%%%%%%%%%%%%%%%%%%%%%%%%%%%%%%%%%%%%%%%%%%%%%%%%%%%
We start at the system with both the TRS and SIS. 
In this case, each pseudospin sector has a pair of the Weyl point at ${\bm k}_c^{(0)}$  and  $-{\bm k}_c^{(0)}$ with the same quasienergy $\phi_c^{(0)}$(=$\pm 1/16$, $\pm 3/16$ in units of $2\pi$). Strictly speaking, they are not the Weyl points, but the Dirac points, because each Weyl cone are doubly degenerate between the two pseudospin sectors.  
By breaking the SIS via the constant gauge field ${\bm A}$, the Weyl points just move in momentum space by $\pm{\bm A}$. Note that the gauge field comes with the minimal coupling ${\bm k}\to {\bm k}-{\bm A}$. 
Since the sign of the gauge field depends on pseudospin, we have four Weyl points  with the same quasienergy $\phi_c^{(0)}$.   Introducing small PSOI lifts the quasienergy and momenta, such that the Weyl points emerge as two pairs by the TRS. Namely, they are found at $(\phi,{\bm k})=(\phi_{c1},{\bm k}_{c1})$, $(\phi_{c1},-{\bm k}_{c1})$, $(\phi_{c2},{\bm k}_{c2})$, and $(\phi_{c2},-{\bm k}_{c2})$. In this way, the slight difference in $\phi_c$ of the Weyl points comes from the PSOI.

The reason why the band touching appears on the $k_x=k_y$ plane is as follows. 
Under the S-matrices of Eqs. (\ref{Eq_S_shift1}) and (\ref{Eq_S_shift2}) with the assumption of $S_1=S_2$, a rotational symmetry holds:
\begin{align}
&{\cal R}^{-1}U_{-k_y,-k_x,-k_z}{\cal R}=U_{\bm k},\\
&{\cal R}= \left(\begin{array}{cccc}
0 & 0 & 0 & \sigma_1\\
0 & 0 & \sigma_1 & 0\\
0 & \sigma_1 & 0 & 0\\
\sigma_1 & 0 & 0 & 0
\end{array}\right).  
\end{align} 
The rotational symmetry is the product of the space inversion and parity, ${\cal R}={\cal S}{\cal P}$, although the latter two symmetries are broken. Combined with the TRS, the quasienergy eigenvalues become symmetric under the exchange of $k_x$ and $k_y$: 
\begin{align}
\phi(k_y,k_x,k_z)=\phi(k_x,k_y,k_z).  
\end{align}
Therefore, the minimum and maximum of the bands appear most typically on the symmetry plane $k_x=k_y$.

\begin{figure}
\centerline{\includegraphics[width=0.3\textwidth]{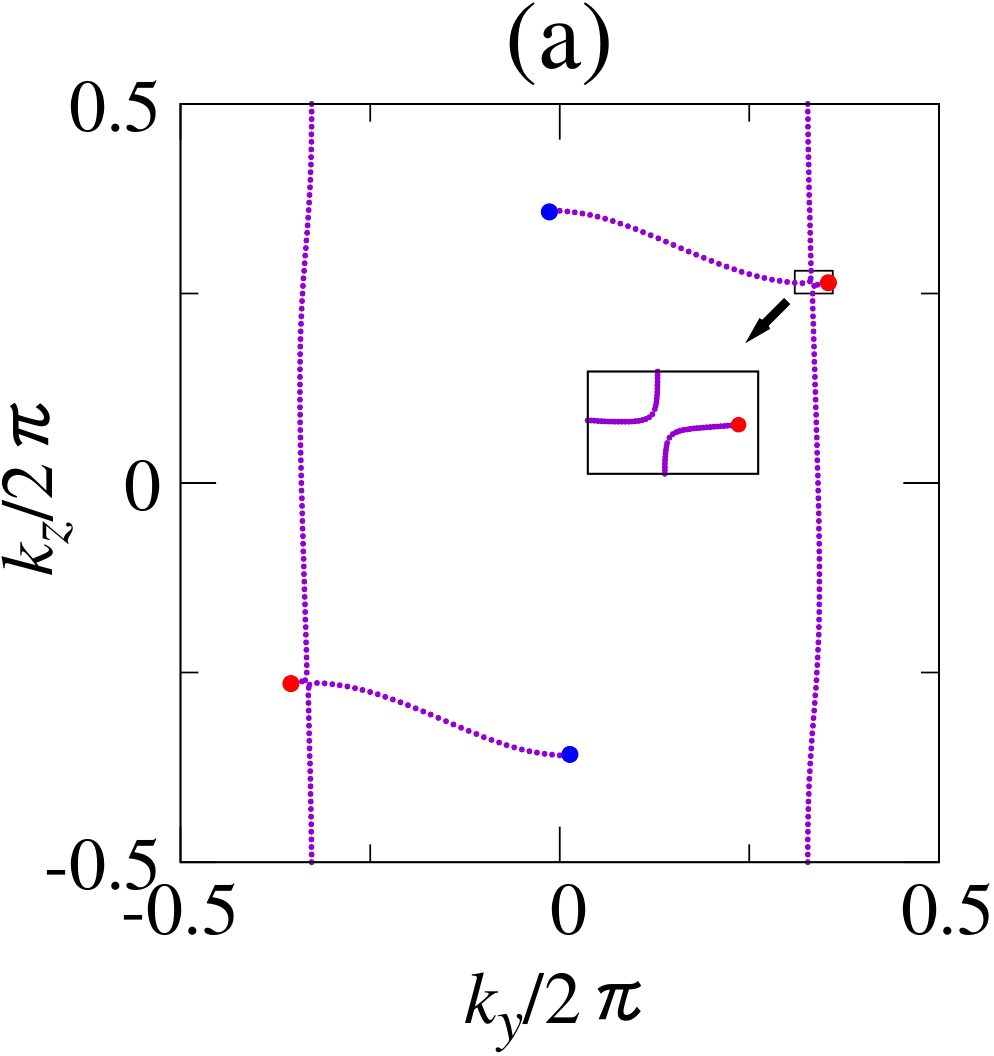}}
\vskip10pt 
\centerline{\includegraphics[width=0.3\textwidth]{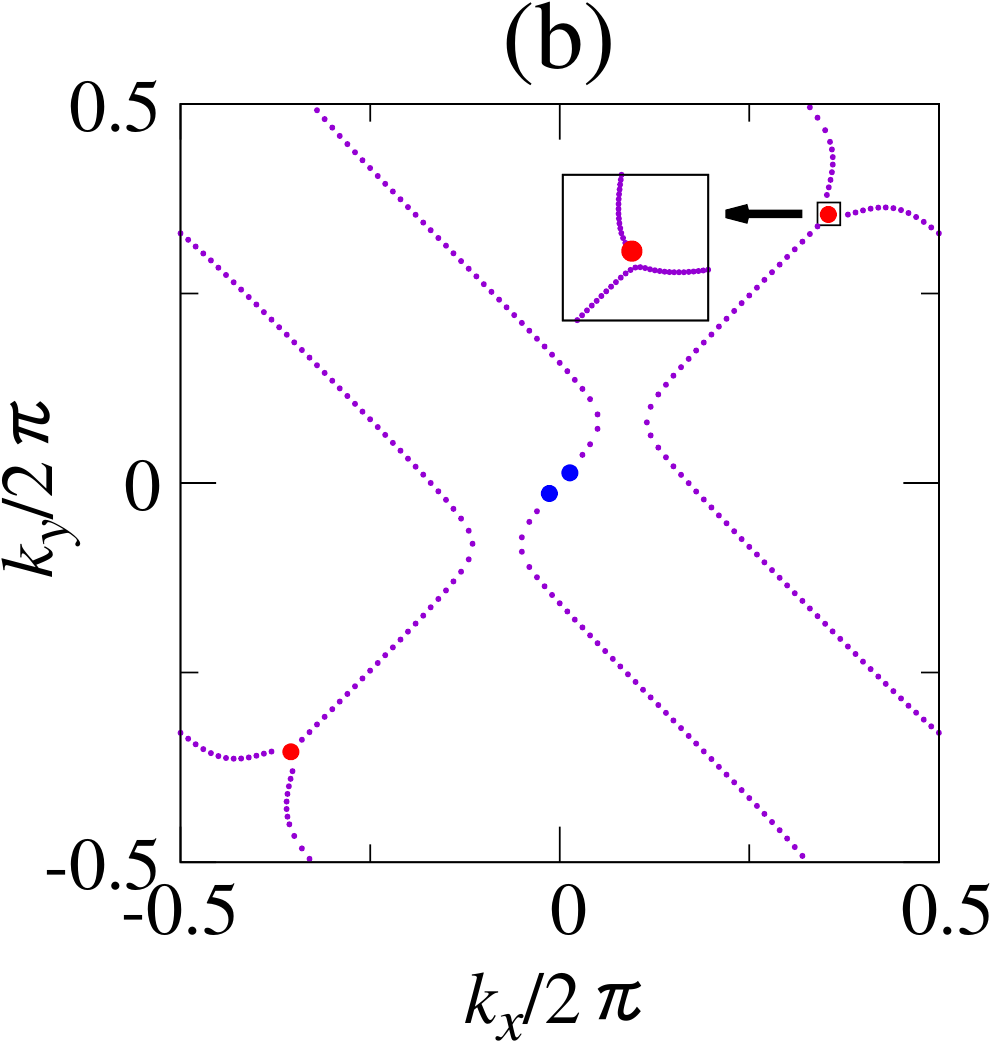}}
%\centerline{\includegraphics[width=0.6\textwidth]{fig9a.pdf}}
%\vskip10pt 
%\centerline{\includegraphics[width=0.6\textwidth]{fig9b.pdf}}
\caption{\label{Fig_soi_weyl_arc} (Color online) Equi-quasienergy contour of the Fermi-arc surface states in the surface normal to the $x$ direction (a) and to the $z$ direction (b). The lower surface is considered. The additional phases  $\varphi_C$ and $\varphi_F$ of the boundary are taken to be zero. 
The following parameters of the S-matrices are employed: $(\alpha_1,\beta_1,\gamma_1,\delta_1)/2\pi=(\alpha_2,\beta_2,\gamma_2,\delta_2)/2\pi=(0,0.02,0.15\exp(-{\rm i}\pi/3),0.02)$,  $(\alpha_3,\beta_3,\gamma_3,\delta_3)/2\pi=(0,0.02,0.31\exp(-{\rm i}\pi/7),0.02)$. The quasienergy value is taken to be $\phi/2\pi=0.3926$, which is the center value between two slightly lifted Weyl-point quasienergies (0.3915 and 0.3936).  The Weyl points of +1 (-1) chirality are indicated by blue (red) dots.    }
\end{figure}

To convince the formation of the Weyl points, we show in Fig. \ref{Fig_soi_weyl_arc}, the equi-quasienergy contour of the surface states at the Weyl-point quasienergy. 
We can see clearly the Fermi arcs connecting two Weyl points having opposite chiralities, irrespective of the surface orientation. 
%%% Fig. 9 %%%%%%%%%%%%%%%%%%%%%%%%%%%%%%%%%%%%%%%%%%%%%%%
%%%%%%%%%%%%%%%%%%%%%%%%%%%%%%%%%%%%%%%%%%%%%%%%%%%%%%
Here, the chirality $\chi$ of the Weyl point at ${\bm k}_c$ is evaluated as 
\begin{align}
&\chi = {\rm sgn}[{\rm det}(b_{ij})],\\
&\psi_{m{\bm k}_c}^\dagger \left( \left. U_{\bm k}^\dagger {\rm i} \frac{\partial}{\partial k_j} U_{\bm k}\right|_{\bm k_c} \right) \psi_{n{\bm k}_c} = \left[a_j\hat{1} + \sum_{l=1}^3 \sigma_l b_{lj}\right]_{mn}, 
\end{align}
where $m$ and $n$ are the band indices of the two degenerate eigenmodes at the Weyl point.

In this way, the FW phase can be preserved in the system with the PSOI by breaking the SIS. If the SIS is unbroken, the PSOI results in the formation of lines nodes around the Weyl points of the unperturbed system (without the PSOI) in the FW phase. By its topological nature, the Weyl points are robust against various perturbations. Therefore, we can expect that the FW phase holds in a wide parameter region of the system with the PSOI.

Although the TRS is difficult to break optically, it is instructive to consider possible effects of broken TRS. In our model, the TRS can be broken, in principle, by introducing a nonreciprocity in the link rings, as shown in the next subsection. However, such a nonreciprocity also breaks the SIS. To isolate possible effects of the TRS breaking, we consider the system with the SIS but without the TRS. We assume that the S-matrices hold Eq. (\ref{Eq_Smat_SIS}) but do not satisfy Eq. (\ref{Eq_Smat_TRS}).  At zero PSOI, the TRS breaking lifts the degeneracy between the clockwise and counter-clockwise modes and causes a shift of the Weyl points in momentum space as in Fig. \ref{Fig_weyl_schematic}. 
%${\bm k}_{c}^{(0)}+{\bm \Delta}_1$ ($\uparrow$), ${\bm k}_{c}^{(0)}+{\bm Delta}_2$ ($\downarrow$), $-{\bm k}_{c}^{(0)}-{\bm \Delta}_1$ ($\uparrow$), and $-{\bm k}_{c}^{(0)}-{\bm \Delta}_2$ ($\downarrow$), with the same quasienergy $\phi_c^{(0)}$. 
The nonzero PSOI results in two pairs of the Weyl points at  $(\phi,{\bm k})=(\phi_{c1},{\bm k}_{c1})$, $(\phi_{c1},-{\bm k}_{c1})$, $(\phi_{c2},{\bm k}_{c2})$, and $(\phi_{c2},-{\bm k}_{c2})$, holding the SIS. 
Note that the chirality pattern is different from that by the SIS breaking. 
The different chirality pattern causes different Fermi-arc configurations.

\subsection{Compatibility of  pseudospin-orbit interaction and synthetic gauge fields}

In the system without the PSOI, the shift of the link rings induces the synthetic gauge field with the U(1) gauge symmetry. However, this gauge symmetry is broken in the system with the PSOI. To implement the U(1) gauge symmetry, we need to have a nonreciprocity in the propagation phase in the link rings, as shown below.

The U(1) gauge symmetry is the symmetry under inclusion of a position-dependent phase $\theta_{\bm r}$: 
\begin{align}
\alpha_{\bm r}\to \alpha_{\bm r}{\rm e}^{{\rm i}\theta_{\bm r}}, \quad 
\alpha'_{\bm r}\to \alpha'_{\bm r}{\rm e}^{{\rm i}\theta_{\bm r}},
\end{align}
To preserve the gauge symmetry, the gauged S-matrix should have the form as  
\begin{align}
S_j(A)=\left(\begin{array}{cc}
S_j^{++} & S_j^{+-}{\rm e}^{-{\rm i}A_{{\bm r};j}} \\
S_j^{-+}{\rm e}^{{\rm i}A_{{\bm r};j}} & S_j^{--}
\end{array}\right).  
\end{align}
The gauge transformation is defined by 
\begin{align}
&A_{{\bm r};j}\to A_{{\bm r};j} + \theta_{{\bm r}+{\bm a}_j}- \theta_{\bm r}.
\end{align}

In the system without the PSOI, the gauge field is implemented with the opposite signs between different pseudospin sectors. As a result, they are subjected to opposite magnetic fields. However, if the same gauge field is applied to different pseudospin sectors, the gauge symmetry is broken.

To maintain the gauge symmetry in the system with the PSOI, we need to have the nonreciprocity.  In Fig. \ref{Fig_setup}(c), we introduced four propagation phases in the link ring. The phases $\phi_1$ and $\phi'_1$ correspond to the propagation phase of the clockwise and counter-clockwise modes, respectively in the upper part of the link ring. The phases  $\phi_2$ and $\phi'_2$ are for the lower part. 
The nonreciprocity implies that $\phi_1\ne \phi'_1$ or  $\phi_2\ne \phi'_2$.

At nodes A and C, we introduce the nodal S-matrices between the site ring and link ring as follows.   
\begin{align}
&\left(\begin{array}{l}
a_{\bm r}\\
a'_{\bm r}\\
\alpha\\
\alpha'
\end{array}\right)=S_{\rm A}\left(\begin{array}{l}
d_{{\bm r}}{\rm e}^{2{\rm i}\phi}\\
e'_{{\bm r}}{\rm e}^{{\rm i}\phi}\\
\gamma{\rm e}^{{\rm i}\phi_2}\\
\gamma'{\rm e}^{{\rm i}\phi'_1}
\end{array}\right),\\
&\left(\begin{array}{l}
c_{{\bm r}+\hat{x}}\\
c'_{{\bm r}+\hat{x}}\\
\gamma\\
\gamma'
\end{array}\right)=S_{\rm C}
\left(\begin{array}{l}
b_{{\bm r}+\hat{x}}{\rm e}^{2{\rm i}\phi}\\
f'_{{\bm r}+\hat{x}}{\rm e}^{{\rm i}\phi}\\
\alpha{\rm e}^{{\rm i}\phi_1}\\
\alpha'{\rm e}^{{\rm i}\phi'_2}
\end{array}\right). 
\end{align}
If we put $\phi_1=\phi'_2$ and $\phi'_1=\phi_2$, we obtain
\begin{align}
&S_1^{++}=S_{\rm A}^{++} +{\rm e}^{{\rm i}(\phi_1+\phi'_1)}S_{\rm A}^{+-} (1-{\rm e}^{{\rm i}(\phi_1+\phi'_1)} S_{\rm C}^{--}S_{\rm A}^{--})^{-1} \nonumber\\
&\hskip50pt \times S_{\rm C}^{--}S_{\rm A}^{-+}, \\
&S_1^{+-}={\rm e}^{{\rm i}\phi'_1}S_{\rm A}^{+-} (1-{\rm e}^{{\rm i}(\phi_1+\phi'_1)} S_{\rm C}^{--}S_{\rm A}^{--})^{-1} S_{\rm C}^{-+},\\
&S_1^{-+}={\rm e}^{{\rm i}\phi_1}S_{\rm C}^{+-} (1-{\rm e}^{{\rm i}(\phi_1+\phi'_1)} S_{\rm A}^{--}S_{\rm C}^{--})^{-1} S_{\rm A}^{-+},\\
&S_1^{--}=S_{\rm C}^{++} +{\rm e}^{ {\rm i}(\phi_1+\phi'_1)}S_{\rm C}^{+-} (1-{\rm e}^{{\rm i}(\phi_1+\phi'_1)} S_{\rm A}^{--}S_{\rm C}^{--})^{-1} \nonumber \\
&\hskip50pt \times S_{\rm A}^{--}S_{\rm C}^{-+}.
\end{align}
If $\phi_1+\phi'_1$ is kept constant while as $\phi_1-\phi'_1$ varies in space, 
we have a synthetic gauge field of 
\begin{align}
A_{{\bm r};1}=\frac{1}{2}(\phi_1-\phi'_1)=-\frac{1}{2}(\phi_2-\phi'_2).  
\end{align}
Therefore, nonzero $A_{{\bm r};1}$ requires the nonreciprocity in the propagation phase, $\phi_1\ne\phi'_1$. 
This form of the hopping S-matrix breaks both the TRS and SIS.

In order to have a synthetic magnetic field whose flux is of order $\pi$, we need to have a huge nonreciprocity of the same order.  In optical systems, it is quite difficult to obtain such a nonreciprocity in a ring-resonator system. Usually, the optical nonreciprocity in the propagation phase involves the magneto-optical effect under the Faraday geometry.  The effect is very small and the nonreciprocity of order $\pi$ is only available by using extremely long optical paths compared to relevant wavelength.    
Thus, it is quite difficult to implement a synthetic gauge field in optical systems with the PSOI.

\section{Conclusion}
 To conclude, we have investigated effects of synthetic gauge fields and PSOI in the stacked 2D ring network lattice. A 3D analogue of the Hofstadter butterfly is obtained under synthetic magnetic fields. The difference in the winding numbers in the gaps above and below the bulk bands is equal to the sum of the Chern numbers of the bulk bands.  This relation is a generalization of the bulk-edge correspondence found in 2D Floquet-Bloch systems.  The PSOI is introduced as a mixing between clockwise and counter-clockwise modes in the rings. It destroys the FTI phase of the system without the PSOI. However, the FW phase can be preserved by breaking the SIS. As a result, multiple Weyl points and Fermi arcs are obtained in the quasienergy spectrum.  The compatibility between the synthetic gauge field and PSOI requires the nonreciprocity in the propagation phase in the link rings.

%\subsection{}
%\subsubsection{}

% \begin{figure}
% \end{figure}

% \begin{table}%[H] add [H] placement to break table across pages
% \begin{ruledtabular}
% Lines of table here ending with \\
% \end{tabular}
% \end{ruledtabular}
% \end{table}

\appendix
\section{Bulk eigenmodes under synthetic magnetic field}
Let us summarize the eigenvalue equation of the bulk modes under synthetic magnetic fields of fractional flux per unit area. For simplicity, we consider two types of the Landau gauge that cover the magnetic fields in the $xy$ and $yz$ planes.

\subsection{ ${\bm B}\cdot \hat{z}=0$}
For the magnetic field parallel in the $xy$ plane, we can choose the following gauge: 
\begin{align}
{\bm A}=(B_yz,-B_xz,0), \quad B_x=2\pi\frac{P_x}{Q_x}, \quad B_y=2\pi\frac{P_y}{Q_y}.  
\end{align}
with coprime integers $P_\mu$ and $Q_\mu$.  
In this case, the translational invariance in the $z$ direction is given by 
$z\to z+2L$, where $L$ is the least common multiple of $Q_x$ and $Q_y$.  This is because the gauge field is included as a phase factor  
$\exp(\pm {\rm i}A_i)$ with $A_1=2\pi P_y/Q_y z$, $A_2=-2\pi P_x/Q_x z$, and $A_3=\pi(P_y/Q_y - P_x/Q_x)z$. 
The eigenvalue equation becomes Eq. (\ref{Eq_Floquet}) with  
\begin{align}
&U_{\bm k}=U_1U_2U_3,\\
&\psi_{\bm k}=(a_0,c_0,a_1,c_1,\dots,a_{4L-1},c_{4L-1})^t,\\
&\alpha_n=\alpha_{{\bm r}=n{\bm s}} \quad (\alpha=a,c),
\end{align}
where $8L\times 8L$ matrices $U_1$, $U_2$, and $U_3$ are given by 
\begin{align}
&U_1={\rm Bdiag}(\tilde{S}_1|_{z=0},\tilde{S}_1|_{z=\frac{1}{2}},\dots,\tilde{S}_1|_{z=2L-\frac{1}{2}}),\\
&U_2={\rm Bdiag}(\tilde{S}_2|_{z=0},\tilde{S}_2|_{z=\frac{1}{2}},\dots,\tilde{S}_2|_{z=2L-\frac{1}{2}}),\\
&(U_3)_{ij}= \scalebox{0.8}{$\displaystyle 
\left\{\begin{array}{ll}
S_3^{++}|_{z=\frac{n}{2}+\frac{1}{4}} \qquad (i,j)=& (1+2n,1+2n)\\
S_3^{+-}|_{z=\frac{n}{2}+\frac{1}{4}}        & (1+2n,4+2n)\\
S_3^{-+}|_{z=\frac{n}{2}+\frac{1}{4}}        & (4+2n,1+2n)\\
S_3^{--}|_{z=\frac{n}{2}+\frac{1}{4}}        & (4+2n,4+2n)\\
 & \quad 0\le n\le 4L-2 \\
S_3^{++}|_{z=2L-\frac{1}{4}} & (8L-1,8L-1)\\
S_3^{+-}|_{z=2L-\frac{1}{4}}{\rm e}^{{\rm i}k_34L} & (8L-1,2)\\
S_3^{-+}|_{z=2L-\frac{1}{4}}{\rm e}^{-{\rm i}k_34L} & (2,8L-1)\\
S_3^{--}|_{z=2L-\frac{1}{4}} & (2,2)\\
0                            & {\rm otherwise}
       \end{array}\right.$}\\
&\tilde{S}_1=\left(\begin{array}{cc}
1 & 0 \\
0 &{\rm e}^{-{\rm i}k_x}
\end{array}\right)S_1
\left(\begin{array}{cc}
0 & 1 \\
{\rm e}^{{\rm i}k_x} & 0
\end{array}\right), \label{Eq_S1t}\\
&\tilde{S}_2=\left(\begin{array}{cc}
1 & 0 \\
0 &{\rm e}^{-{\rm i}k_y}
\end{array}\right)S_2
\left(\begin{array}{cc}
1 & 0 \\
0 &{\rm e}^{{\rm i}k_y}
\end{array}\right).\label{Eq_S2t}
\end{align}

The quasienergy eigenvalues satisfy the extra-symmetry:  
\begin{align}
\phi\left(k_x-\pi\frac{P_y}{Q_y},k_y+\pi\frac{P_x}{Q_x},k_z\right)=\phi(k_x,k_y,k_z),
\end{align}
in addition to the symmetries by definition: 
\begin{align}
&\phi(k_x+2\pi,k_y,k_z)=\phi(k_x,k_y,k_z),\\
&\phi(k_x,k_y+2\pi,k_z)=\phi(k_x,k_y,k_z),\\
&\phi\left(k_x,k_y,k_z+\pi\frac{1}{L}\right)=\phi(k_x,k_y,k_z),
\end{align} 
The former symmetry comes from the invariance of the eigenvalue equation under 
\begin{align}
&(k_x,k_y,k_z)\to \left(k_x-\pi\frac{P_y}{Q_y},k_y+\pi\frac{P_x}{Q_x},k_z\right),\\
&\alpha_n\to \alpha_{n+1}{\rm e}^{{\rm i}\lambda_n},\\
&\lambda_{n+1}-\lambda_{n}=\frac{\pi}{2}\left(\frac{P_x}{Q_x}-\frac{P_y}{Q_y}\right). 
\end{align}

\subsection{${\bm B}\cdot \hat{x}=0$}
For the magnetic field in the $yz$ plane, we can choose the following gauge:
\begin{align}
{\bm A}=(0,B_zx,-B_yx), \quad B_y=2\pi\frac{P_y}{Q_y}, \quad B_z=2\pi\frac{P_z}{Q_z}.  
\end{align}  
The translational invariance in the $x$ direction becomes $x\to x+2L$, where $L$ is the least common multiple of $Q_y$ and $Q_z$. 

In this case, the unitary matrix to be diagonalized becomes
\begin{align}
&U_{\bm k}=U_1U_2U_3,\\
&\psi_{\bm k}=(a_0^{(1)},c_0^{(1)},a_0^{(2)},c_0^{(2)},\dots,a_{2L-1}^{(1)},c_{2L-1}^{(1)},a_{2L-1}^{(2)},c_{2L-1}^{(2)})^t,\\
&\alpha_n^{(1)}=\alpha_{{\bm r}=n\hat{x}}, \quad \alpha_n^{(2)}=\alpha_{{\bm r}=n\hat{x}+{\bm s}},  
\end{align}
where $8L\times 8L$ matrices  $U_1$, $U_2$, and $U_3$ are given by 
\begin{align}
&(U_1)_{ij}=\scalebox{0.8}{$\displaystyle 
\left\{\begin{array}{ll}
S_1^{++}\qquad (i,j)=&(1+2n,2+2n)\\
S_1^{+-}       &(1+2n,5+2n)\\
S_1^{-+}       &(6+2n,2+2n)\\
S_1^{--}       &(6+2n,5+2n)\\
               &\quad 0\le n\le 4L-3\\
S_1^{++}       &(8L-3,8L-2),(8L-1,8L)\\
S_1^{+-}{\rm e}^{{\rm i}k_x2L}  &(8L-3,1),(8L-1,3)\\
S_1^{-+}{\rm e}^{-{\rm i}k_x2L}  & (2,8L-2),(4,8L)\\
S_1^{--}      &(2,1),(4,3)\\
0         &{\rm otherwise}
      \end{array}\right.$},\\
&U_2={\rm Bdiag}(\tilde{S}_2|_{x=0},\tilde{S}_2|_{x=\frac{1}{2}},\dots,\tilde{S}_2|_{x=2L-\frac{1}{2}}),\\
&(U_3)_{ij}=\scalebox{0.8}{$\displaystyle
\left\{\begin{array}{ll}
S_3^{++}|_{x=n+\frac{1}{4}} \qquad (i,j)=&(1+4n,1+4n)\\
S_3^{+-}|_{x=n+\frac{1}{4}} & (1+4n,4+4n)\\
S_3^{-+}|_{x=n+\frac{1}{4}} & (4+4n,1+4n)\\
S_3^{--}|_{x=n+\frac{1}{4}} & (4+4n,4+4n)\\
                            & \quad  0\le n \le 2L-1\\
S_3^{++}|_{x=n+\frac{3}{4}} &  (3+4n,3+4n)\\
S_3^{+-}|_{x=n+\frac{3}{4}}{\rm e}^{ {\rm i}(k_y+k_z)} &  (3+4n,6+4n)\\
S_3^{-+}|_{x=n+\frac{3}{4}}{\rm e}^{-{\rm i}(k_y+k_z)} &  (6+4n,3+4n)\\
S_3^{--}|_{x=n+\frac{3}{4}} &  (6+4n,6+4n)\\ 
                           & \quad  0\le n \le 2L-2\\
S_3^{++}|_{x=2L-\frac{1}{4}} &  (8L-1,8L-1)\\
S_3^{+-}|_{x=2L-\frac{1}{4}}{\rm e}^{ {\rm i}(k_x2L+k_y+k_z)} &  (8L-1,2)\\
S_3^{-+}|_{x=2L-\frac{1}{4}}{\rm e}^{-{\rm i}(k_x2L+k_y+k_z)} &  (2,8L-1)\\
S_3^{--}|_{x=2L-\frac{1}{4}} &  (2,2)\\ 
0 & {\rm otherwise}               
\end{array}\right.$}.
\end{align}

The quasienergy eigenvalues satisfy the following symmetry relation:
\begin{align}
\phi\left(k_x,k_y-\pi\frac{P_z}{Q_z},k_z+\pi\frac{P_y}{Q_y}\right)=\phi(k_x,k_y,k_z),
\end{align}
in addition to the symmetries by definition: 
\begin{align}
&\phi\left(k_x+\pi\frac{1}{L},k_y,k_z\right)=\phi(k_x,k_y,k_z),\\
&\phi(k_x,k_y+2\pi,k_z)=\phi(k_x,k_y,k_z),\\
&\phi(k_x,k_y,k_z+2\pi)=\phi(k_x,k_y,k_z).
\end{align} 
The former symmetry comes from the invariance of the eigenvalue equation under 
\begin{align}
&(k_x,k_y,k_z) \to \left(k_x,k_y-\pi\frac{P_z}{Q_z},k_z+\pi\frac{P_y}{Q_y}\right),\\
&\alpha_n^{(1)}\to \alpha_{n}^{(2)}{\rm e}^{{\rm i}\lambda_1},\\
&\alpha_n^{(2)}\to \alpha_{n+1}^{(1)}{\rm e}^{{\rm i}\lambda_2},\\
&\lambda_{2}-\lambda_{1}=k_y+k_z-\frac{\pi}{2}\left(\frac{P_z}{Q_z}-\frac{P_y}{Q_y}\right). 
\end{align}

\section{Slab S-matrices under synthetic magnetic fields}

In this section, we present a construction of the slab S-matrix for finite-thick  systems.   
To obtain the slab S-matrix, we just need a monolayer S-matrix. 
The layer-doubling scheme \cite{Pendry-LEED-book} allows us to calculate the S-matrix of $N$-layer thick slab from the monolayer S-matrix.

We assume that the slab has finite-thickness in the direction parallel to ${\bm n}$ and infinite extent in the direction normal to ${\bm n}$. We denote the slab S-matrix as $S_{\mu;N}$ for ${\bm n}=\hat{\mu}$ and $N$-layer thick slab.

\subsection{${\bm n}=\hat{z}$ and ${\bm B}\cdot\hat{z}=0$}   
For the slab system along the $z$ direction under the synthetic magnetic field in the $xy$ plane, we employ the Landau gauge of ${\bm A}=(B_yz,-B_xz,0)$ that preserves the translational invariance in the $x$ and $y$ directions. 
The slab S-matrix $S_{z;N}$ is introduced as 
\begin{align}
&\left(\begin{array}{c}
a_N \\
e_0
\end{array}\right)=S_{z;N}
\left(\begin{array}{c}
a_0 \\
e_N
\end{array}\right),\\
&\alpha_n = \alpha_{{\bm r}=n{\bm s}}.  
\end{align}
It can be constructed from the monolayer S-matrix $S_{z;1}({\bm r})$ defined by 
\begin{align}
&\left(\begin{array}{c}
a_{{\bm r}+{\bm s}} \\
e_{\bm r}
\end{array}\right)=S_{z;1}({\bm r})
\left(\begin{array}{c}
a_{\bm r} \\
e_{{\bm r}+{\bm s}}
\end{array}\right). 
\end{align}
From the definition of the hopping S-matrices Eqs. (\ref{Eq_S1}-\ref{Eq_S3}), together with Eqs. (\ref{Eq_S1t}) and (\ref{Eq_S2t}), we obtain  
\begin{align}
&S_{z;1}({\bm r})=\Xi_3\left(z+\frac{1}{4}\right) \otimes \Xi_{12}\left(z+\frac{1}{2}\right),\label{Eq_LD1}
\end{align}
with 
\begin{align}
&\Xi_{12}(z)=\tilde{S}_1\tilde{S}_2\sigma_1{\rm e}^{3{\rm i}\phi},\\
&\Xi_3(z)=\sigma_1 S_3{\rm e}^{{\rm i}\phi}.
\end{align}
%Here, ${S_j|_z}$ refers to the $j$-th component of the S-matrix [Eqs. (\ref{Eq_S1}-\ref{Eq_S3})] whose gauge-field term is given by  ${\bm A}=(B_yz,-B_xz,0)$. 
Here, symbol $\otimes$ represents  the ``S-matrix product'' defined by 
\begin{align}
&(S_L\otimes S_R)^{++}=S_R^{++}(1-S_L^{+-}S_R^{-+})^{-1}S_L^{++},\\
&(S_L\otimes S_R)^{+-}=S_R^{++}(1-S_L^{+-}S_R^{-+})^{-1}S_L^{+-}S_R^{--} + S_R^{+-},\\
&(S_L\otimes S_R)^{-+}=S_L^{--}(1-S_R^{-+}S_L^{+-})^{-1}S_R^{-+}S_L^{++} + S_L^{-+},\\
&(S_L\otimes S_R)^{--}=S_L^{--}(1-S_R^{-+}S_L^{+-})^{-1}S_R^{--},
\end{align}
which satisfies the associativity:
\begin{align}
(S_1\otimes S_2)\otimes S_3 = S_1\otimes (S_2\otimes S_3). 
\end{align}
The slab S-matrix is obtained from the monolayer S-matrix as 
\begin{align}
S_{z;N}=S_{z;1}({\bm 0})\otimes S_{z;1}({\bm s})\otimes \cdots \otimes S_{z;1}((N-1){\bm s}). \label{Eq_LDNz}
\end{align}

To obtain the surface states, the boundary condition at the upper and lower surfaces of the slab is required. 
It is generally expressed as 
\begin{align}
&e_N = T_z^{({\rm u})}a_N,\\
&a_0 = T_z^{({\rm l})}e_0,  
\end{align}
with unitary $T_z^{({\rm u})}$ and $T_z^{({\rm l})}$. 
We impose the following boundary conditions with additional phases at nodes E and F of the site rings in the upper and lower surfaces, respectively:
\begin{align}
&e_N={\rm e}^{{\rm i}(\phi+\varphi_E)}a_N,\\
&f_0={\rm e}^{{\rm i}(\phi+\varphi_F)}c_0.
\end{align}
This boundary condition results in 
\begin{align}
&T_z^{({\rm u})} ={\rm e}^{  {\rm i}(\phi+\varphi_E)},\label{Eq_TzN}\\
&T_z^{({\rm l})} = \Xi_{12}^{+-}(0) + \Xi_{12}^{++}(0){\rm e}^{{\rm i}(\phi+\varphi_F)} \nonumber \\
&\hskip50pt \times [1-{\rm e}^{{\rm i}(\phi+\varphi_F)}\Xi_{12}^{-+}(0)]^{-1}\Xi_{12}^{--}(0).\label{Eq_Tz0} 
\end{align}

\subsection{${\bm n}=\hat{z}$ and ${\bm B}\propto\hat{z}$}   
For the slab along the $z$ direction under the magnetic field in the $z$ direction, we employ the gauge field of  ${\bm A}=(0,B_zx,0)$. 
The slab S-matrix $S_{z;N}$ is defined by 
\begin{align}
&\left(\begin{array}{c}
A_N \\
E_0
\end{array}\right)=S_{z;N}
\left(\begin{array}{c}
A_0 \\
E_N
\end{array}\right),\\
&A_n=(a_{n{\bm s}},a_{n{\bm s}+\hat{x}},\dots,a_{n{\bm s}+(2Q_z-1)\hat{x}})^t.
\end{align}
It can be constructed from the monolayer S-matrix $S_{z;1}$ defined by 
\begin{align}
&\left(\begin{array}{c}
A_{{\bm r}+{\bm s}} \\
E_{\bm r}
\end{array}\right)=S_{z;1}({\bm r})
\left(\begin{array}{c}
A_{\bm r} \\
E_{{\bm r}+{\bm s}}
\end{array}\right),\\
&A_{\bm r}=(a_{{\bm r}},a_{{\bm r}+\hat{x}},\dots,a_{{\bm r}+(2Q_z-1)\hat{x}})^t.  
\end{align}

The monolayer S-matrix is given by Eq. (\ref{Eq_LD1}), where  
\begin{widetext}
\begin{align}
&\Xi_{12}(z)=\Xi_1\Xi_2(z),\\
&[\Xi_1]_{ij}={\rm e}^{2{\rm i}\phi}\scalebox{0.8}{$\displaystyle
\left\{\begin{array}{ll}
S_1^{++} \qquad \qquad \qquad (i,j)=&(n+1,n+1+2Q_z)\\
S_1^{--}              &(n+1+2Q_z,n+1)\\
                      &\qquad  0\le n \le 2Q_z-1\\ 
S_1^{+-}              &(n+1,n+2)\\
S_1^{-+}              &(n+2+2Q_z,n+1+2Q_z)\\
                      &\qquad  0\le n \le 2Q_z-2\\
S_1^{+-}{\rm e}^{ik_x2Q_z}  &(2Q_z,1)\\
S_1^{-+}{\rm e}^{-ik_x2Q_z} &(1+2Q_z,4Q_z)\\
0                     &{\rm otherwise}
\end{array}\right.$},\\
&[\Xi_2(z)]_{ij}={\rm e}^{{\rm i}\phi}\scalebox{0.8}{$\displaystyle
\left\{\begin{array}{ll}
S_2^{++}|_{x=n+z} \qquad (i,j)=&(n+1,n+1+2Q_z)\\
S_2^{+-}|_{x=n+z}{\rm e}^{{\rm i}k_y}      &(n+1,n+1)\\
S_2^{-+}|_{x=n+z}{\rm e}^{-{\rm i}k_y}     &(n+1+2Q_z,n+1+2Q_z)\\
S_2^{--}|_{x=n+z}              &(n+1+2Q_z,n+1)\\
                                     & \qquad 0\le n\le 2Q_z-1\\
0 & {\rm otherwise}
\end{array}\right.$},\\
&[\Xi_3(z)]_{ij}={\rm e}^{{\rm i}\phi}\scalebox{0.8}{$\displaystyle
\left\{\begin{array}{ll}
S_3^{-+}|_{x=n+z} \qquad (i,j)=& (n+1,n+1)\\
S_3^{--}|_{x=n+z} & (n+1,n+1+2Q_z)\\
S_3^{++}|_{x=n+z} & (n+1+2Q_z,n+1)\\
S_3^{+-}|_{x=n+z} & (n+1+2Q_z,n+1+2Q_z)\\
& \qquad 0\le n\le 2Q_z-1\\
0 & {\rm otherwise}
\end{array}\right.$}. 
\end{align}
\end{widetext}
The slab S-matrix is obtained by Eq. (\ref{Eq_LDNz}).

The boundary condition at the upper and lower surfaces of the slab is generally expressed as 
\begin{align}
&E_N = T_z^{({\rm u})}A_N,\\
&A_0 = T_z^{({\rm l})}E_0.  
\end{align}
If we impose the following boundary condition: 
\begin{align}
&E_N={\rm e}^{{\rm i}(\phi+\varphi_E)}A_N,\\
&F_0={\rm e}^{{\rm i}(\phi+\varphi_F)}C_0,
\end{align}
with additional phase $\varphi_{E(F)}$ at node E(F) of the upper (lower) surface, 
we obtain the same forms of $T_z^{({\rm u})}$ and $T_z^{({\rm l})}$ as Eqs. (\ref{Eq_TzN}) and (\ref{Eq_Tz0}), respectively.

\subsection{${\bm n}=\hat{x}$ and ${\bm B}\cdot\hat{x}=0$}  
For the slab along the $x$ direction under the magnetic field in the $yz$ plane, we employ the gauge ${\bm A}=(0,B_zx,-B_yx)$. The translational invariance in the $y$ and $z$ direction is preserved. 
The slab S-matrix is defined by  
\begin{align}
&\left(\begin{array}{c}
f_N^{(1)} \\
c_N^{(2)}\\
b_0^{(1)}\\
b_0^{(2)}
\end{array}\right)=S_{x;N}
\left(\begin{array}{c}
f_0^{(1)} \\
c_0^{(2)}\\
b_N^{(1)}\\
b_N^{(2)}
\end{array}\right),\\
&\alpha_n^{(1)}=\alpha_{n\hat{x}},\quad 
\alpha_n^{(2)}=\alpha_{n\hat{x}+{\bm s}}. 
\end{align}
The monolayer S-matrix defined by 
\begin{align}
&\left(\begin{array}{c}
f_{{\bm r}+\hat{x}} \\
c_{{\bm r}+\hat{x}+{\bm s}}\\
b_{{\bm r}}\\
b_{{\bm r}+{\bm s}}
\end{array}\right)=S_{x;1}({\bm r})
\left(\begin{array}{c}
f_{{\bm r}}\\
c_{{\bm r}+{\bm s}}\\
b_{{\bm r}+\hat{x}}\\
b_{{\bm r}+\hat{x}+{\bm s}}
\end{array}\right). 
\end{align}
has the following expression:
\begin{widetext}
\begin{align}
&S_{x;1}({\bm r})= T_4\Lambda' + T_3\Lambda(1-T_1\Lambda)^{-1}T_2\Lambda',\\
&T_1 = \scalebox{0.8}{$\displaystyle \left(\begin{array}{cccccccc}
     0 & 0 & S_1^{++} & 0 & 0 & 0 & 0 & 0\\
     0 & 0 & S_1^{-+} & 0 & 0 & 0 & 0 & 0\\
     0 & 0 & 0 & S_2^{-+}|_x{\rm e}^{-{\rm i}k_y} & 0 & 0 & 0 & 0\\
     S_3^{++}|_{x+\frac{1}{4}} & 0 & 0 & 0 & 0 & 0 & 0 & 0\\
     0 & 0 & 0 & 0 & 0 & S_1^{++} & 0 & 0\\
     0 & 0 & 0 & 0 & 0 & 0 & S_2^{-+}|_{x+\frac{1}{2}}{\rm e}^{-{\rm i}k_y} & S_2^{--}|_{x+\frac{1}{2}}\\
     0 & S_3^{+-}|_{x+\frac{3}{4}}{\rm e}^{{\rm i}(k_y+k_z)} & 0 & 0 & S_3^{++}|_{x+\frac{3}{4}} & 0 & 0 & 0\\
     S_3^{-+}|_{x+\frac{1}{4}} & 0 & 0 & 0 & 0 & 0 & 0 & 0\\
	   \end{array}\right)$},\\
&T_2 =\scalebox{0.8}{$\displaystyle \left(\begin{array}{cccc}
 0 & 0 & S_1^{+-} & 0\\
 0 & 0 & S_1^{--} & 0\\
 S_2^{--}|_x & 0 & 0 & 0\\
 0 & S_3^{+-}|_{x+\frac{1}{4}} & 0 & 0\\
 0 & 0 & 0 & S_1^{+-}\\
 0 & 0 & 0 & 0\\
 0 & 0 & 0 & 0\\
 0 & S_3^{--}|_{x+\frac{1}{4}} & 0 & 0
\end{array}\right)$},\\
&T_3 =\scalebox{0.8}{$\displaystyle \left(\begin{array}{cccccccc}
0 &S_3^{--}|_{x+\frac{3}{4}} & 0 & 0 & S_3^{-+}|_{x+\frac{3}{4}}{\rm e}^{-{\rm i}(k_y+k_z)} & 0 & 0 & 0\\
0 & 0 & 0 & 0 & 0 & S_1^{++} & 0 & 0\\
0 & 0 & 0 & S_2^{++}|_x & 0 & 0 & 0 & 0\\
0 & 0 & 0 & 0 & 0 & 0 & S_2^{++}|_{x+\frac{1}{2}} & S_2^{+-}|_{x+\frac{1}{2}}{\rm e}^{{\rm i}k_y}\\
\end{array}\right)$},\\
&T_4 = \scalebox{0.8}{$\displaystyle \left(\begin{array}{cccc}
0 & 0 & 0 & 0\\
0 & 0 & 0 & S_1^{+-}\\
S_2^{+-}|_x{\rm e}^{{\rm i}k_y}& 0 & 0 & 0\\
0 & 0 & 0 & 0\\
\end{array}\right)$},\\
&\Lambda={\rm diag}({\rm e}^{{\rm i}\phi},{\rm e}^{{\rm i}\phi},{\rm e}^{2{\rm i}\phi},{\rm e}^{{\rm i}\phi},{\rm e}^{{\rm i}\phi},{\rm e}^{2{\rm i}\phi},{\rm e}^{{\rm i}\phi},{\rm e}^{{\rm i}\phi}),\\
&\Lambda'={\rm diag}({\rm e}^{{\rm i}\phi},{\rm e}^{{\rm i}\phi},{\rm e}^{2{\rm i}\phi},{\rm e}^{2{\rm i}\phi}). 
\end{align}
\end{widetext}
The slab S-matrix is obtained from the monolayer S-matrix as
\begin{align}
S_{x;N}=S_{x;1}({\bm 0})\otimes S_{x;1}(\hat{x})\otimes \cdots \otimes S_{x;1}((N-1)\hat{x}). \label{Eq_LDNx}
\end{align}

\clearpage

The boundary condition at the slab surfaces is generally expressed as 
\begin{align}
&\left(\begin{array}{l}
b_N^{(1)} \\
b_N^{(2)} 
\end{array}\right)=T_x^{({\rm u})}
\left(\begin{array}{l}
f_N^{(1)} \\
c_N^{(2)} 
\end{array}\right),\\
&\left(\begin{array}{l}
f_0^{(1)} \\
c_0^{(2)} 
\end{array}\right)=T_x^{({\rm l})}
\left(\begin{array}{l}
b_0^{(1)} \\
b_0^{(2)} 
\end{array}\right).
\end{align}
If we impose additional phase at nodes A, C, E, and F on the boundary surfaces as 
\begin{align}
&a_N^{(1)}=d_N^{(1)}{\rm e}^{ {\rm i}(2\phi+\varphi_A)},\\
&a_N^{(2)}=d_N^{(2)}{\rm e}^{ {\rm i}(2\phi+\varphi_A)},\\
&e_N^{(2)}=a_N^{(2)}{\rm e}^{ {\rm i}(\phi + \varphi_E)},\\
&c_0^{(1)}=b_0^{(1)}{\rm e}^{ {\rm i}(2\phi+\varphi_C)},\\
&c_0^{(2)}=b_0^{(2)}{\rm e}^{ {\rm i}(2\phi+\varphi_C)},\\
&f_0^{(1)}=c_0^{(1)}{\rm e}^{ {\rm i}(\phi + \varphi_F)},
\end{align}    
we obtain 
\begin{widetext}
\begin{align}
&T_x^{({\rm u})}={\rm e}^{{\rm i}\phi}V_4 +{\rm e}^{  2{\rm i}\phi}V_3\Lambda(1-{\rm e}^{{\rm i}\phi}V_1\Lambda)^{-1}V_2,\\
&T_x^{({\rm l})}={\rm diag}[{\rm e}^{{\rm i}(3\phi+\varphi_C+\varphi_F)},{\rm e}^{{\rm i}(2\phi+\varphi_C)}],\\
&V_1=\left(\begin{array}{cccc}
0 & S_2^{-+}|_{x=N}{\rm e}^{-{\rm i}k_y} & 0 & 0 \\
S_3^{++}|_{x=N+\frac{1}{4}} & 0 & 0 & 0\\
0 & 0 & S_2^{-+}|_{x=N+\frac{1}{2}}{\rm e}^{-{\rm i}k_y} & 
 S_2^{--}|_{x=N+\frac{1}{2}}\\
S_3^{-+}|_{x=N+\frac{1}{4}} & 0 & 0 & 0
\end{array}\right),\\
&V_2=\left(\begin{array}{cc}
S_2^{--}|_{x=N} & 0  \\
0 & S_3^{+-}|_{x=N+\frac{1}{4}} \\
0 & 0 \\
0 & S_3^{--}|_{x=N+\frac{1}{4}} 
\end{array}\right),\\
&V_3=\left(\begin{array}{cccc}
0 & S_2^{++}|_{x=N} & 0 & 0 \\
0 & 0 & S_2^{++}|_{x=N+\frac{1}{2}} & S_2^{+-}|_{x=N+\frac{1}{2}}{\rm e}^{{\rm i}k_y} 
\end{array}\right),\\
&V_4=\left(\begin{array}{cc}
S_2^{+-}|_{x=N}{\rm e}^{{\rm i}k_y} & 0  \\
0 & 0
\end{array}\right),\\
&\Lambda={\rm diag}({\rm e}^{{\rm i}(2\phi+\varphi_A)},1,{\rm e}^{{\rm i}(3\phi+\varphi_A+\varphi_E)},1).
\end{align}
\end{widetext}

\subsection{${\bm n}=\hat{x}$ and ${\bm B}\propto\hat{x}$}   
For the slab along the $x$ direction under the magnetic field in the $x$ direction, we employ the gauge ${\bm A}=(0,-B_xz,0)$.  
The slab S-matrix is defined by 
\begin{align}
&\left(\begin{array}{c}
F_N^{(1)} \\
C_N^{(2)}\\
B_0^{(1)}\\
B_0^{(2)}
\end{array}\right)=S_{x;N}
\left(\begin{array}{c}
F_0^{(1)} \\
C_0^{(2)}\\
B_N^{(1)}\\
B_N^{(2)}
\end{array}\right),\\
&B_n^{(1)}=(b_{n\hat{x}},b_{n\hat{x}+\hat{z}},\dots,b_{n\hat{x}+(2Q_x-1)\hat{z}})^t,\\
&B_n^{(2)}=(b_{{\bm s}+n\hat{x}},b_{{\bm s}+n\hat{x}+\hat{z}},\dots,b_{{\bm s}+ n\hat{x}+(2Q_x-1)\hat{z}})^t. 
\end{align}
The monolayer S-matrix is defined by 
\begin{align}
&\left(\begin{array}{c}
F_{{\bm r}+\hat{x}} \\
C_{{\bm r}+{\bm s}+\hat{x}}\\
B_{{\bm r}}\\
B_{{\bm r}+{\bm s}}
\end{array}\right)=S_{x;1}
\left(\begin{array}{c}
F_{{\bm r}} \\
C_{{\bm r}+{\bm s}}\\
B_{{\bm r}+\hat{x}}\\
B_{{\bm r}+{\bm s}+\hat{x}}
\end{array}\right),\\
&B_{\bm r}=(b_{{\bm r}},b_{{\bm r}+\hat{z}},\dots,b_{{\bm r}+(2Q_x-1)\hat{z}})^t.
\end{align}
It is given by 
\begin{widetext}
\begin{align}
&S_{x;1}= T_4\Lambda' + T_3\Lambda(1-T_1\Lambda)^{-1}T_2\Lambda',\\
&[T_1]_{ij}=\scalebox{0.8}{$\displaystyle  
\left\{\begin{array}{ll}
S_1^{++} \qquad (i,j)=& (n+1,n+1+4Q_x)\\
S_1^{-+}             & (n+1+2Q_x,n+1+4Q_x) \\
S_2^{-+}|_{z=n}{\rm e}^{  -{\rm i}k_y} & (n+1+4Q_x,n+1+6Q_x)\\
S_3^{++}|_{z=n+\frac{1}{4}} & (n+1+6Q_x,n+1)\\
S_1^{++}                 & (n+1+8Q_x,n+1+10Q_x)\\
S_2^{-+}|_{z=n+\frac{1}{2}}{\rm e}^{-{\rm i}k_y} &(n+1+10Q_x,n+1+12Q_x)\\
S_2^{--}|_{z=n+\frac{1}{2}} &(n+1+10Q_x,n+1+14Q_x)\\
S_3^{++}|_{z=n+\frac{3}{4}}{\rm e}^{{\rm i}k_y} &(n+1+12Q_x,n+1+8Q_x)\\
S_3^{-+}|_{z=n+\frac{1}{4}} & (n+1+14Q_x,n+1)\\
                         & \quad 0\le n \le 2Q_x-1 \\ 
S_3^{+-}|_{z=n+\frac{3}{4}}{\rm e}^{{\rm i}k_y} &(n+1+12Q_x,n+2+2Q_x)\\
                         & \quad 0\le n \le 2Q_x-2 \\
S_3^{+-}|_{z=n+\frac{3}{4}}{\rm e}^{{\rm i}(k_y+k_z2Q_x)} & (14Q_x,1+2Q_x)\\     
0   & {\rm otherwise}                          
\end{array}\right.$},\\
&[T_2]_{ij}=\scalebox{0.8}{$\displaystyle 
\left\{\begin{array}{ll}
S_1^{+-} \qquad (i,j)=& (n+1,n+1+4Q_x)\\
S_1^{--}              & (n+1+2Q_x,n+1+4Q_x)\\
S_2^{--}|_{z=n}         & (n+1+4Q_x,n+1)\\
S_3^{+-}|_{z=n+\frac{1}{4}}         & (n+1+6Q_x,n+1+2Q_x)\\
S_1^{+-}              & (n+1+8Q_x,n+1+6Q_x)\\
S_3^{--}|_{z=n+\frac{1}{4}}         & (n+1+14Q_x,n+1+2Q_x)\\
                      & \quad 0\le n \le 2Q_x-1 \\ 
0   & {\rm otherwise}          
\end{array}\right.$},\\
&[T_3]_{ij}=\scalebox{0.8}{$\displaystyle 
\left\{\begin{array}{ll}
S_1^{++}\qquad (i,j)=&(n+1+2Q_x,n+1+10Q_x)\\
S_2^{++}|_{z=n} &(n+1+4Q_x,n+1+6Q_x)\\
S_2^{++}|_{z=n+\frac{1}{2}} &(n+1+6Q_x,n+1+12Q_x)\\
S_2^{+-}|_{z=n+\frac{1}{2}}{\rm e}^{{\rm i}k_y} &(n+1+6Q_x,n+1+14Q_x)\\
S_3^{--}|_{z=n+\frac{3}{4}}{\rm e}^{-{\rm i}k_y} & (n+1,n+1+8Q_x)\\
            & \quad 0\le n \le 2Q_x-1\\
S_3^{-+}|_{z=n+\frac{3}{4}}{\rm e}^{-{\rm i}k_y} & (n+2,n+1+2Q_x)\\
            & \quad 0\le n \le 2Q_x-2\\
S_3^{--}|_{z=n+\frac{3}{4}}{\rm e}^{-{\rm i}(k_y+k_z2Q_x)} & (1,4Q_x)\\
0  & {\rm otherwise} 
\end{array}\right.$},\\
&[T_4]_{ij}=\scalebox{0.8}{$\displaystyle \left\{\begin{array}{ll}
S_1^{+-} \qquad (i,j)=&(n+1+2Q_x,n+1+6Q_x)\\
S_2^{+-}|_{z=n}{\rm e}^{ {\rm i}k_y} & (n+1+4Q_x,n+1)\\
 & \quad 0\le n \le 2Q_x-1\\
0 & {\rm otherwise} 
\end{array}\right.$},
\end{align}
\begin{align}
&\Lambda={\rm diag}(\Lambda_1,\Lambda_2,\dots,\Lambda_{16Q_x})\\\
&\Lambda_j=\scalebox{0.8}{$\displaystyle
\left\{\begin{array}{ll}
{\rm e}^{{\rm i}\phi} \quad j=&n+1,n+1+2Q_x,n+1+6Q_x,n+1+8Q_x,n+1+12Q_x,n+1+14Q_x  \\
{\rm e}^{2{\rm i}\phi} &n+1+4Q_x,n+1+10Q_x \\
& \quad 0\le n \le 2Q_x-1
\end{array}
\right.$},\\
&\Lambda'={\rm diag}(\Lambda'_1,\Lambda'_2,\dots,\Lambda'_{8Q_x})\\
&\Lambda'_j=\scalebox{0.8}{$\displaystyle
\left\{\begin{array}{ll}
{\rm e}^{{\rm i}\phi} & 1\le j \le 4Q_x  \\
{\rm e}^{2{\rm i}\phi} &4Q_x+1\le j \le 8Q_x 
\end{array}
\right.$}.  
\end{align}
\end{widetext}
The slab S-matrix is given by 
\begin{align}
S_{x;N}=\stackrel{N}{\overbrace{S_{x;1}\otimes \cdots \otimes S_{x;1}}}. 
\end{align}

The boundary condition at the slab surfaces is generally expressed as 
\begin{align}
&\left(\begin{array}{l}
B_N^{(1)} \\
B_N^{(2)} 
\end{array}\right)=T_x^{({\rm u})}
\left(\begin{array}{l}
F_N^{(1)} \\
C_N^{(2)} 
\end{array}\right),\\
&\left(\begin{array}{l}
F_0^{(1)} \\
C_0^{(2)} 
\end{array}\right)=T_x^{({\rm l})}
\left(\begin{array}{l}
B_0^{(1)} \\
B_0^{(2)} 
\end{array}\right).
\end{align}
If we impose additional phase at nodes A, C, E, and F on the boundary surfaces as 
\begin{align}
&A_N^{(1)}=D_N^{(1)}{\rm e}^{ {\rm i} (2\phi+\varphi_A)},\\
&A_N^{(2)}=D_N^{(2)}{\rm e}^{ {\rm i} (2\phi+\varphi_A)},\\
&E_N^{(2)}=A_N^{(2)}{\rm e}^{ {\rm i} (\phi + \varphi_E)},\\
&C_0^{(1)}=B_0^{(1)}{\rm e}^{ {\rm i} (2\phi+\varphi_C)},\\
&C_0^{(2)}=B_0^{(2)}{\rm e}^{ {\rm i} (2\phi+\varphi_C)},\\
&F_0^{(1)}=C_0^{(1)}{\rm e}^{ {\rm i} (\phi + \varphi_F)},
\end{align}    
we obtain 
\begin{widetext}
\begin{align}
&[T_x^{({\rm u})}]_{ij}=\scalebox{0.8}{$\displaystyle
\left\{\begin{array}{ll}
T_n^{++} \qquad (i,j)=&(n+1,n+1) \\
T_n^{+-} &(n+1,n+1+2Q_x) \\
T_n^{-+} &(n+1+2Q_x,n+1) \\
T_n^{--} &(n+1+2Q_x,n+1+2Q_x) \\
         & \qquad 0\le n \le 2Q_x-1 \\
0 & {\rm otherwise}
\end{array}\right.$},\\
&T_x^{({\rm l})}={\rm diag}[\stackrel{2Q_x}{\overbrace{{\rm e}^{{\rm i}(3\phi+\varphi_C+\varphi_F)},\dots}},\stackrel{2Q_x}{\overbrace{{\rm e}^{{\rm i}(2\phi+\varphi_C)},\dots}}],\\
&T_n={\rm e}^{{\rm i}\phi}V_4 +{\rm e}^{  2{\rm i}\phi} V_3\Lambda(1-{\rm e}^{{\rm i}\phi}V_1\Lambda)^{-1}V_2,\\
&V_1=\left(\begin{array}{cccc}
0 & S_2^{-+}|_{z=n}{\rm e}^{  -ik_y}& 0 & 0 \\
S_3^{++}|_{z=n+\frac{1}{4}} & 0 & 0 & 0\\
0 & 0 & S_2^{-+}|_{z=n+\frac{1}{2}}{\rm e}^{  -{\rm i}k_y}& S_2^{--}|_{z=n+\frac{1}{2}}\\
S_3^{-+}|_{z=n+\frac{1}{4}} & 0 & 0 & 0
\end{array}\right),\\
&V_2=\left(\begin{array}{cc}
S_2^{--}|_{z=n} & 0  \\
0 & S_3^{+-}|_{z=n+\frac{1}{4}} \\
0 & 0 \\
0 & S_3^{--}|_{z=n+\frac{1}{4}} 
\end{array}\right),\\
&V_3=\left(\begin{array}{cccc}
0 & S_2^{++}|_{z=n} & 0 & 0 \\
0 & 0 & S_2^{++}|_{z=n+\frac{1}{2}} & S_2^{+-}|_{z=n+\frac{1}{2}}{\rm e}^{  {\rm i}k_y}
\end{array}\right),\\
&V_4=\left(\begin{array}{cc}
S_2^{+-}|_{z=n}{\rm e}^{  {\rm i}k_y}& 0  \\
0 & 0
\end{array}\right),\\
&\Lambda={\rm diag}[{\rm e}^{{\rm i}(2\phi+\varphi_A)},1,{\rm e}^{{\rm i}(3\phi+\varphi_A+\varphi_E)},1].
\end{align}
\end{widetext}

\section{Slab S-matrix for the system with the pseudospin-orbit interaction}

\subsection{${\bm n}\propto \hat{z}$}
For a slab system having finite thickness in the $z$ direction, the slab S-matrix is defined by 
\begin{align}
&\left(\begin{array}{c}
a_N \\
b'_N\\
e_0\\
e'_0
\end{array}\right)=S_{z;N}
\left(\begin{array}{c}
a_0 \\
b'_0\\
e_N\\
e'_N
\end{array}\right),\\
&\alpha_n=\alpha_{{\bm r}=n{\bm s}}, \quad  \alpha'_n=\alpha_{{\bm r}=n{\bm s}}
\end{align}
The monolayer S-matrix $S_{z;1}$ is obtained from Eqs. (\ref{Eq_S1_PSOI}-\ref{Eq_S3_PSOI}) together with Eqs. (\ref{Eq_S1t_PSOI}) and (\ref{Eq_S2t_PSOI}) as 

\clearpage

\begin{align}
&S_{z;1}=\Xi_3\otimes \Xi_{12},\\
&\Xi_{12}=
\left(\begin{array}{cccc}
1&0 & 0& 0\\
0&0 & 1& 0\\
0&1 & 0& 0\\
0&0 & 0& 1
\end{array}\right) Q
\left(\begin{array}{cccc}
0&0 & 1& 0\\
1&0 & 0& 0\\
0&0 & 0& 1\\
0&1 & 0& 0
\end{array}\right),\\
&\Xi_3=\left(\begin{array}{cc}
	0&\hat{1} \\
	 \hat{1}&0\end{array}\right)S_3{\rm e}^{{\rm i}\phi},\\
&Q=\left(\begin{array}{cc}
Q^{++} & Q^{+-}\\
Q^{-+} & Q^{--}
\end{array}\right),\\
&Q^{++}={\rm e}^{4{\rm i}\phi} \tilde{S}_1^{++}(\hat{1}-{\rm e}^{4{\rm i}\phi}\tilde{S}_2^{+-}\tilde{S}_1^{-+} )^{-1}\tilde{S}_2^{++},\\
&Q^{+-}={\rm e}^{6{\rm i}\phi} \tilde{S}_1^{++}(\hat{1}-{\rm e}^{4{\rm i}\phi}\tilde{S}_2^{+-}\tilde{S}_1^{-+} )^{-1}\tilde{S}_2^{+-}\tilde{S}_1^{--} \nonumber \\
& \hskip50pt +{\rm e}^{  2{\rm i}\phi} \tilde{S}_1^{+-},\\
&Q^{-+}={\rm e}^{6{\rm i}\phi} \tilde{S}_2^{--}(\hat{1}-{\rm e}^{4{\rm i}\phi}\tilde{S}_1^{-+}\tilde{S}_2^{+-} )^{-1}\tilde{S}_1^{-+}\tilde{S}_2^{++} \nonumber \\
& \hskip50pt +{\rm e}^{  2{\rm i}\phi} \tilde{S}_2^{-+},\\
&Q^{--}={\rm e}^{4{\rm i}\phi} \tilde{S}_2^{--}(\hat{1}-{\rm e}^{4{\rm i}\phi}\tilde{S}_1^{-+}\tilde{S}_2^{+-} )^{-1}\tilde{S}_1^{--}.
\end{align}
The slab S-matrix is obtained from the monolayer S-matrix as 
\begin{align}
S_{z;N}=\stackrel{N}{\overbrace{S_{z;1}\otimes \cdots \otimes S_{z;1}}}. 
\end{align}

The boundary condition at the slab surfaces is generally expressed as  
\begin{align}
&\left(\begin{array}{l}
e_N \\
e'_N
\end{array}\right)=T_z^{({\rm u})}
\left(\begin{array}{l}
a_N \\
b'_N
\end{array}\right),\\
&\left(\begin{array}{l}
a_0 \\
b'_0
\end{array}\right)=T_z^{({\rm l})}
\left(\begin{array}{l}
e_0 \\
e'_0
\end{array}\right).
\end{align}
We impose the additional phases at node E and F of the boundary surfaces as 
\begin{align}
&e_N=a_N{\rm e}^{{\rm i}(\phi+\varphi_E)},\\
&e'_N=b'_N{\rm e}^{{\rm i}(\phi+\varphi_E)},\\
&f_0 = c_0{\rm e}^{{\rm i}(\phi+\varphi_F)},\\
&f'_0 = d'_0{\rm e}^{{\rm i}(\phi+\varphi_F)}. 
\end{align}
This condition results in  
\begin{align}
&T_z^{({\rm u})}={\rm e}^{{\rm i}(\phi+\varphi_E)},\\
&T_z^{({\rm l})}=\Xi_{12}^{+-} + \Xi_{12}^{++}{\rm e}^{{\rm i}(\phi+\varphi_F)}[1-{\rm e}^{{\rm i}(\phi+\varphi_F)}\Xi_{12}^{-+}]^{-1}\Xi_{12}^{--}. 
\end{align}

\subsection{${\bm n}\propto \hat{x}$}

For the slab system having finite thickness in the $x$ direction, we define the slab S-matrix as 
\begin{align}
&\left(\begin{array}{l}
f_N^{(1)} \\
c_N^{(1)}{}' \\
c_N^{(2)} \\
c_N^{(2)}{}' \\
b_0^{(1)} \\
d_0^{(1)}{}' \\
b_0^{(2)} \\
f_0^{(2)}{}'
\end{array}\right)=S_{x;N}
\left(\begin{array}{l}
f_0^{(1)} \\
c_0^{(1)}{}' \\
c_0^{(2)} \\
c_0^{(2)}{}' \\
b_N^{(1)} \\
d_N^{(1)}{}' \\
b_N^{(2)} \\
f_N^{(2)}{}'
\end{array}\right),\\
&\alpha_n^{(1)}=\alpha_{{\bm r}+n\hat{x}}, \quad   
\alpha_n^{(2)}=\alpha_{{\bm r}+n\hat{x}+{\bm s}}. 
\end{align}

At $N=1$, the S-matrix is symbolically given by 
\begin{widetext}
\begin{align}
&S_{x;1}=T_4\Lambda' + T_3\Lambda (1-T_1\Lambda)^{-1} T_2\Lambda',\\
&T_1=\scalebox{0.7}{$\displaystyle \left(\begin{array}{cccccccccccccccc}
0 & 0 & S_1^{11}& 0 & 0 & 0 & S_1^{12}& S_1^{14}& 0 & 0 & 0 & 0 & 0 & 0 & 0 & 0\\
0 & 0 & S_1^{31}& 0 & 0 & 0 & S_1^{32}& S_1^{34}& 0 & 0 & 0 & 0 & 0 & 0 & 0 & 0\\
0 & 0 & 0 & S_2^{31}P_y^{-1}& S_2^{34} & 0 & 0 & 0 & 0 & 0 & 0 & 0 & 0 & 0 & 0 & 0\\
S_3^{11} & 0 & 0 & 0 & 0 & S_3^{12} & 0 & 0 & 0 & 0 & 0 & 0 & 0 & 0 & S_3^{14} & 0\\
0 & 0 & S_1^{21}& 0 & 0 & 0 & S_1^{22}& S_1^{24}& 0 & 0 & 0 & 0 & 0 & 0 & 0 & 0\\
0 & 0 & 0 & S_2^{21}& S_2^{24}P_y & 0 & 0 & 0 & 0 & 0 & 0 & 0 & 0 & 0 & 0 & 0\\
S_3^{21} & 0 & 0 & 0 & 0 & S_3^{22} & 0 & 0 & 0 & 0 & 0 & 0 & 0 & 0 & S_3^{24} & 0\\
0 & S_3^{43} & 0 & 0 & 0 & 0 & 0 & 0 & S_3^{41}P_{yz}^{-1} & 0 & 0 & 0 & 0 & S_3^{42}P_{yz}^{-1} & 0 & 0\\ 
0 & 0 & 0 & 0 & 0 & 0 & 0 & 0 & 0 & S_x^{11} & 0 & 0 & 0 & 0 & 0 & S_1^{12}\\
0 & 0 & 0 & 0 & 0 & 0 & 0 & 0 & 0 & 0 & S_2^{31}P_y^{-1} & S_2^{33} & S_2^{34} & 0 & 0 & 0\\
0 & S_3^{13}P_{yz} & 0 & 0 & 0 & 0 & 0 & 0 & S_3^{11} & 0 & 0 & 0 & 0 & S_3^{12} & 0 & 0\\
S_3^{31} & 0 & 0 & 0 & 0 & S_3^{32} & 0 & 0 & 0 & 0 & 0 & 0 & 0 & 0 & S_3^{34} & 0\\
0 & 0 & 0 & 0 & 0 & 0 & 0 & 0 & 0 & S_1^{21} & 0 & 0 & 0 & 0 & 0 & S_1^{22}\\
0 & 0 & 0 & 0 & 0 & 0 & 0 & 0 & 0 & 0 & S_2^{21} & S_2^{23}P_y & S_2^{24}P_y & 0 & 0 & 0\\
0 & 0 & 0 & 0 & 0 & 0 & 0 & 0 & 0 & 0 & S_2^{41}P_y^{-1} & S_2^{43} & S_2^{44} & 0 & 0 & 0\\
0 & S_3^{23}P_{yz} & 0 & 0 & 0 & 0 & 0 & 0 & S_3^{21} & 0 & 0 & 0 & 0 & S_3^{22} & 0 & 0
	 \end{array}\right)$},\nonumber \\
& \\
&T_2=\scalebox{0.7}{$\displaystyle \left(\begin{array}{cccccccc}
0 & 0 & 0 & 0 & S_1^{13} & 0 & 0 & 0 \\
0 & 0 & 0 & 0 & S_1^{33} & 0 & 0 & 0 \\
S_2^{33} & S_2^{32}P_y^{-1} & 0 & 0 & 0 & 0 & 0 & 0 \\
0 & 0 & S_3^{13} & 0 & 0 & 0 & 0 & 0 \\
0 & 0 & 0 & 0 & S_1^{23} & 0 & 0 & 0 \\
S_2^{23}P_y & S_2^{22} & 0 & 0 & 0 & 0 & 0 & 0 \\
0 & 0 & S_3^{23}  & 0 & 0 & 0 & 0 & 0 \\
0 & 0 & 0 & 0 & 0 & S_3^{44} & 0 & 0 \\
0 & 0 & 0 & 0 & 0 & 0 & S_1^{13} & S_1^{14} \\
0 & 0 & 0 & S_2^{32}P_y^{-1} & 0 & 0 & 0 & 0 \\
0 & 0 & 0 & 0 & 0 & S_3^{14}P_{yz} & 0 & 0 \\
0 & 0 & S_3^{33}  & 0 & 0 & 0 & 0 & 0 \\
0 & 0 & 0 & 0 & 0 & 0 & S_1^{23} & S_1^{24} \\
0 & 0 & 0 & S_2^{22} & 0 & 0 & 0 & 0 \\
0 & 0 & 0 & S_2^{42}P_y^{-1}  & 0 & 0 & 0 & 0 \\
0 & 0 & 0 & 0 & 0 & S_3^{24}P_{yz} & 0 & 0 
	 \end{array}\right)$},
\end{align}
\begin{align}
&T_3=\scalebox{0.7}{$\displaystyle \left(\begin{array}{cccccccccccccccc}
 0 & S_3^{33} & 0 & 0 & 0 & 0 & 0 & 0 & S_3^{31}P_{yz}^{-1}& 0 & 0 & 0 & 0 & S_3^{32}P_{yz}^{-1} & 0 & 0\\
 0 & 0 & S_1^{41} & 0 & 0 & 0 & S_1^{42} & S_1^{44} & 0 & 0 & 0 & 0 & 0 & 0 & 0 & 0\\
 0 & 0 & 0 & 0 & 0 & 0 & 0 & 0 & 0 & S_1^{31} & 0 & 0 & 0 & 0 & 0 & S_1^{32}\\
 0 & 0 & 0 & 0 & 0 & 0 & 0 & 0 & 0 & S_1^{41} & 0 & 0 & 0 & 0 & 0 & S_1^{42}\\
 0 & 0 & 0 & S_2^{11} & S_2^{14}P_y & 0 & 0 & 0 & 0 & 0 & 0 & 0 & 0 & 0 & 0 & 0\\
 0 & 0 & 0 & S_2^{41}P_y^{-1} & S_2^{44} & 0 & 0 & 0 & 0 & 0 & 0 & 0 & 0 & 0 & 0 & 0\\
 0 & 0 & 0 & 0 & 0 & 0 & 0 & 0 & 0 & 0 & S_2^{11} & S_2^{13}P_y & S_2^{14}P_y & 0 & 0 & 0\\
 S_3^{41} & 0 & 0 & 0 & 0 & S_3^{42} & 0 & 0 & 0 & 0 & 0 & 0 & 0 & 0 & S_3^{44} & 0
	 \end{array}\right)$},\nonumber \\
& \\
&T_4=\scalebox{0.7}{$\displaystyle \left(\begin{array}{cccccccc}
   0 & 0 & 0 & 0 & 0 & S_3^{34} & 0 & 0\\
   0 & 0 & 0 & 0 & S_1^{43} & 0 & 0 & 0\\
   0 & 0 & 0 & 0 & 0 & 0 & S_1^{33} & S_1^{34}\\
   0 & 0 & 0 & 0 & 0 & 0 & S_1^{43} & S_1^{44}\\
   S_2^{13}P_y & S_2^{12} & 0 & 0 & 0 & 0 & 0 & 0\\
   S_2^{43} & S_2^{42}P_y^{-1} & 0 & 0 & 0 & 0 & 0 & 0\\
   0 & 0 & 0 & S_2^{12} & 0 & 0 & 0 & 0\\
   0 & 0 & S_3^{43} & 0 & 0 & 0 & 0 & 0
	 \end{array}\right)$},\\
&\Lambda={\rm diag}({\rm e}^{{\rm i}\phi},{\rm e}^{{\rm i}\phi},{\rm e}^{2{\rm i}\phi},{\rm e}^{{\rm i}\phi},{\rm e}^{2{\rm i}\phi},{\rm e}^{{\rm i}\phi},{\rm e}^{{\rm i}\phi},{\rm e}^{{\rm i}\phi},{\rm e}^{{\rm i}\phi},{\rm e}^{2{\rm i}\phi},{\rm e}^{{\rm i}\phi},{\rm e}^{{\rm i}\phi},{\rm e}^{2{\rm i}\phi},{\rm e}^{{\rm i}\phi},{\rm e}^{{\rm i}\phi},{\rm e}^{{\rm i}\phi}),\\
&\Lambda'={\rm diag}({\rm e}^{{\rm i}\phi},{\rm e}^{2{\rm i}\phi},{\rm e}^{{\rm i}\phi},{\rm e}^{2{\rm i}\phi},{\rm e}^{2{\rm i}\phi},{\rm e}^{{\rm i}\phi},{\rm e}^{2{\rm i}\phi},{\rm e}^{{\rm i}\phi}),\\
&P_y={\rm e}^{{\rm i}k_y}, \quad P_{yz}={\rm e}^{{\rm i}(k_y+k_z)}
\end{align}
\end{widetext}
The slab S-matrix is obtained from the monolayer S-matrix as 
\begin{align}
S_{x;N}=\stackrel{N}{\overbrace{S_{x;1}\otimes \cdots \otimes S_{x;1}}}. 
\end{align}

The boundary condition at the slab surfaces  are expressed  as 
\begin{align}
&\left(\begin{array}{l}
b_N^{(1)}\\
d_N^{(1)}{}'\\
b_N^{(2)}\\
f_N^{(2)}{}'
\end{array}\right)=T_x^{({\rm u})}
\left(\begin{array}{l}
f_N^{(1)}\\
c_N^{(1)}{}'\\
c_N^{(2)}\\
c_N^{(2)}{}'
\end{array}\right),\\
&\left(\begin{array}{l}
f_0^{(1)}\\
c_0^{(1)}{}'\\
c_0^{(2)}\\
c_0^{(2)}{}'
\end{array}\right)=T_x^{({\rm u})}
\left(\begin{array}{l}
b_0^{(1)}\\
d_0^{(1)}{}'\\
b_0^{(2)}\\
f_0^{(2)}{}'
\end{array}\right). 
\end{align}
If we impose the additional phases on the boundary surfaces as  
\begin{align}
&a_N^{(1)}=d_N^{(1)}{\rm e}^{{\rm i}(2\phi+\varphi_A)},\\
&a_N^{(1)}{}'=e_N^{(1)}{}'{\rm e}^{{\rm i}(\phi+\varphi_A)},\\
&a_N^{(2)}=d_N^{(2)}{\rm e}^{{\rm i}(2\phi+\varphi_A)},\\
&a_N^{(2)}{}'=e_N^{(2)}{}'{\rm e}^{{\rm i}(\phi+\varphi_A)},\\
&e_N^{(2)}=a_N^{(2)}{\rm e}^{{\rm i}(\phi+\varphi_E)},\\
&e_N^{(2)}{}'=b_N^{(2)}{}'{\rm e}^{{\rm i}(\phi+\varphi_E)},\\
&c_0^{(1)}=b_0^{(1)}{\rm e}^{{\rm i}(2\phi+\varphi_C)},\\
&c_0^{(1)}{}'=f_0^{(1)}{}'{\rm e}^{{\rm i}(\phi+\varphi_C)},\\
&f_0^{(1)}=c_0^{(1)}{\rm e}^{{\rm i}(\phi+\varphi_F)},\\
&f_0^{(1)}{}'=d_0^{(1)}{}'{\rm e}^{{\rm i}(\phi+\varphi_F)},\\
&c_0^{(2)}=b_0^{(2)}{\rm e}^{{\rm i}(2\phi+\varphi_C)},\\
&c_0^{(2)}{}'=f_0^{(2)}{}'{\rm e}^{{\rm i}(\phi+\varphi_C)},
\end{align}
we obtain 
\begin{widetext}
\begin{align}
&T_x^{({\rm u})}=V_4\Lambda'+V_3\Lambda (1-V_1\Lambda)^{-1}V_2\Lambda',\\
&T_x^{({\rm l})}={\rm diag}\left[{\rm e}^{{\rm i}(3\phi+\varphi_C+\varphi_F)},{\rm e}^{{\rm i}(2\phi+\varphi_C+\varphi_F)},{\rm e}^{{\rm i}(2\phi+\varphi_C)},{\rm e}^{{\rm i}(\phi+\varphi_C)}\right],\\
&V_1=\left(\begin{array}{cccccccc}
0 & S_2^{31}P_y^{-1} & 0 & S_2^{34} & 0 & 0 & 0 & 0\\
S_3^{11} & 0 & S_3^{12} & 0 & 0 & 0 & 0 & S_3^{14} \\
0 & S_2^{21} & 0 & S_2^{24}P_y & 0 & 0 & 0 & 0\\
S_3^{21} & 0 & S_3^{22} & 0 & 0 & 0 & 0 & S_3^{24} \\
0 & 0 & 0 & 0 & S_2^{31}P_y^{-1} & S_2^{33} & S_2^{34} & 0\\ 
S_3^{31} & 0 & S_3^{32} & 0 & 0 & 0 & 0 & S_3^{34} \\
0 & 0 & 0 & 0 & S_2^{21} & S_2^{23}P_y & S_2^{24}P_y & 0\\ 
0 & 0 & 0 & 0 & S_2^{41}P_y^{-1} & S_2^{43} & S_2^{44} & 0
	 \end{array}\right),\\
&V_2=\left(\begin{array}{cccc}
S_2^{33} & S_2^{32}P_y^{-1} & 0 & 0\\
0 & 0 & S_3^{13} & 0 \\
S_2^{23}P_y & S_2^{22} & 0 & 0\\
0 & 0 & S_3^{23} & 0 \\
0 & 0 & 0 & S_2^{32}P_y^{-1}\\
0 & 0 & S_3^{33} & 0 \\
0 & 0 & 0 & S_2^{22}\\
0 & 0 & 0 & S_2^{42}P_y^{-1}
\end{array}\right),\\
&V_3=\left(\begin{array}{cccccccc}
0 & S_2^{11} & 0 &  S_2^{13}P_y & 0 & 0 & 0 & 0\\
0 & S_2^{41}P_y^{-1} & 0 &  S_2^{43} & 0 & 0 & 0 & 0\\
0 & 0 & 0 & 0 & S_2^{11} & S_2^{13}P_y & S_2^{14}P_y & 0\\
S_3^{41} & 0 & S_3^{42} & 0 & 0 & 0 & 0 & S_3^{44}
\end{array}\right),\\
&V_4=\left(\begin{array}{cccc}
S_2^{13}P_y & S_2^{12} & 0 & 0\\
S_2^{43} & S_2^{42}P_y^{-1} & 0 & 0\\
0 & 0 & 0 & S_2^{12}\\
0 & 0 & S_3^{43} & 0
\end{array}\right),\\
&\Lambda={\rm diag}\left[{\rm e}^{{\rm i}(3\phi+\varphi_A)},{\rm e}^{{\rm i}\phi},{\rm e}^{{\rm i}\phi},{\rm e}^{{\rm i}(3\phi+\varphi_A)},{\rm e}^{{\rm i}(4\phi+\varphi_A+\varphi_E)},{\rm e}^{{\rm i}\phi},{\rm e}^{{\rm i}(4\phi+\varphi_A+\varphi_E)},{\rm e}^{{\rm i}\phi}\right],\\
&\Lambda'={\rm diag}\left[{\rm e}^{{\rm i}\phi},{\rm e}^{2{\rm i}\phi},{\rm e}^{{\rm i}\phi},{\rm e}^{2{\rm i}\phi}\right].
\end{align}
\end{widetext}

\begin{acknowledgments}
%\ack  %IOP 
This work was partially supported by JSPS KAKENHI Grant No. 26390013. 
\end{acknowledgments}

%\section*{References}
%\bibliography{../../Database/mydata}


\end{document}